\documentclass[11pt,fleqn,a4paper]{article}

\usepackage[mathscr]{eucal}
\usepackage{amsmath,amssymb,amsthm}
\usepackage{graphicx}
\usepackage{subcaption}
\usepackage{cite}
\usepackage{hyperref}

\hypersetup{colorlinks, linkcolor=blue, citecolor=blue, urlcolor=blue}

\allowdisplaybreaks

\newcommand{\ve}{\varepsilon}

{\theoremstyle{definition}

\newtheorem{remark}{Remark}
\newtheorem*{remark*}{Remark}
}

\newcommand{\todo}[1][\null]{\ensuremath{\clubsuit}}

\newcommand{\noprint}[1]{}
\newcommand{\checked}[1][\null]{\ensuremath{\boldsymbol{\surd}}}
\newcommand{\pdl}[2]{\frac{\partial #1}{\partial #2}}

\begin{document}

\par\noindent {\LARGE\bf
Stochastic domain decomposition for the solution\\ of the two-dimensional magnetotelluric problem
\par}

{\vspace{4mm}\par\noindent {\bf Alexander Bihlo$^\dag$, Colin G.\ Farquharson$^\ddag$, Ronald D.\ Haynes$^\dag$\\ and J.\ Concepci\'{o}n Loredo-Osti$^\dag$
} \par\vspace{2mm}\par}

{\vspace{2mm}\par\noindent {\it
$^{\dag}$~Department of Mathematics and Statistics, Memorial University of Newfoundland,\\ St.\ John's (NL) A1C 5S7, Canada
}}
{\vspace{2mm}\par\noindent {\it
$^{\ddag}$~Department of Earth Sciences, Memorial University of Newfoundland,\\ St.\ John's (NL) A1B 3X5, Canada
}}

{\vspace{2mm}\par\noindent {\it
\textup{E-mail:} abihlo@mun.ca, cgfarquh@mun.ca, rhaynes@mun.ca, jcloredoosti@mun.ca
}\par}

{\vspace{5mm}\par\noindent\hspace*{5mm}\parbox{150mm}{\small
 Stochastic domain decomposition is proposed as a novel method for solving the two-dimensional Maxwell's equations as used in the magnetotelluric method. The stochastic form of the exact solution of Maxwell's equations is evaluated using Monte-Carlo methods taking into consideration that the domain may be divided into neighboring sub-domains. These sub-domains can be naturally chosen by splitting the sub-surface domain into regions of constant (or at least continuous) conductivity. The solution over each sub-domain is obtained by solving Maxwell's equations in the strong form. The sub-domain solver used for this purpose is a meshless method resting on radial basis function based finite differences. The method is demonstrated by solving a number of classical magnetotelluric problems, including the quarter-space problem, the block-in-half-space problem and the triangle-in-half-space problem.
}\par\vspace{2mm}}

\section{Introduction}

The magnetotelluric method is a standard remote sensing method for inferring the Earth sub- surface electric structure by measuring, at the Earth's surface, the electro-magnetic fields arising from electric currents induced in the sub-surface by naturally occurring time variations of the Earth's magnetic field. Due to its potential of probing the sub-surface conductivity structure up to several hundred kilometres, the magnetotelluric method has become a standard technique for exploration surveys aiming at locating mineral and hydrocarbon resources, and for investigating the structure and composition of the Earth's crust and upper mantle~\cite{chav12a,vozo91a}.

Linking the data obtained from field surveys to the conductivity structure in the ground requires a numerical solution of Maxwell's equations. Several techniques have been proposed for this purpose, including finite difference, finite volume and finite element solvers~\cite{weis12a,zhda97a}. In this paper we propose another method suitable for the numerical evaluation of Maxwell's equations, based on {\it stochastic domain decomposition}~\cite{aceb05a}, that is particularly suited to efficient computation via parallelization and that has the potential to handle arbitrary topography and realistically complex geological interfaces.

Stochastic domain decomposition is a relatively recent domain decomposition method. In traditional (deterministic) domain decomposition one generally splits the physical domain into sub-domains and alternately (or in parallel) solves the given differential equation over each sub-domain. Proper interface conditions between the sub-domains ensure convergence of the domain decomposition procedure to the global solution over the entire domain using an iteration procedure. The rate of convergence strongly depends on what type of interface conditions are chosen, see e.g.~\cite{quar99a}, and finding the most optimized interface conditions is usually a challenging task.

In contrast to traditional domain decomposition, stochastic domain decomposition does not require iteration. The main requirement for the applicability of stochastic domain decomposition is that the partial differential equation under consideration possesses a stochastic representation of its exact solution. This is always the case for linear elliptic boundary value problems and linear parabolic initial--boundary value problems~\cite{kara91a}. Certain nonlinear differential equations also allow for a stochastic representation of their exact solution, see e.g.~\cite{aceb10a}.

The probabilistic form of the exact solution of a differential equation can be evaluated numerically using Monte-Carlo methods. While Monte-Carlo methods are known to converge notoriously slowly and hence only become competitive for higher-dimensional problems~\cite{pres07Ay}, the situation is different in the stochastic domain decomposition framework. Here, one evaluates the stochastic representation of the exact solution only on the interfaces between the sub-domains. Thus, rather than computing the solution of the global problem using the stochastic technique at every point, only interface solutions have to be computed. The solution over each individual sub-domain can then be obtained using deterministic methods. Moreover, since the stochastic solution reproduces the exact solution up to the numerical error (consisting of a time-stepping error, the boundary hitting error and the Monte-Carlo error~\cite{aceb05a}), no iteration is required for the domain decomposition technique to converge. In addition, once the interface solutions are obtained, the sub-domain solutions can be computed over all the sub-domains simultaneously and in parallel. Stochastic domain decomposition is thus particularly suited to massively parallel computing architectures. Stochastic domain decomposition has been used to solve physical partial differential equations in ~\cite{aceb05a,aceb07a,aceb10a} and for the generation of moving meshes in partial differential based grid generators in~\cite{bihl14a,bihl15b}.

In this paper we apply the stochastic domain decomposition method to the two-dimensional magnetotelluric problem, solving the two-dimensional Maxwell's equation in the time-frequency domain. The main challenge in applying the method to Maxwell's equations is the presence of conductivity jumps. The two-dimensional Maxwell's equations, as used in the magnetotelluric method, are a system of differential equations with discontinuous coefficients. This adds another layer of complexity as most work connecting boundary value problems and stochastic calculus has been done for equations with continuous coefficients, see e.g.~\cite{mile12Ay}. However, the discontinuity in the conductivity allows for a natural splitting in sub-domains, namely those where the conductivity is constant (or at least continuous). This enables one to solve Maxwell's equations in the \emph{strong form} on each of the sub-domains, which is the route that we will pursue in this paper. For a recent exposition on a deterministic domain decomposition method for the three-dimensional time-dependent Maxwell's equations, see~\cite{dole14a}.

Since the stochastic form of the exact solution of Maxwell's equations can be evaluated at arbitrary points, and in realistic sub-surface models the conductivity jumps can have arbitrary shape, it is natural to use a deterministic sub-domain solver that can handle a variety of interface layouts as well. This makes so-called \emph{meshfree methods} a natural choice. The use of meshfree methods in geophysics is relatively recent. The magnetotelluric problem has been considered quite recently in this light in~\cite{witt14a}, although there the authors used the Maxwell's equation in the weak form. This requires one to use high order numerical integration which can be avoided if meshless methods in the strong form are invoked. In the present paper, we will use radial basis function based finite differences (RBF-FD). This is a prominent meshless method~\cite{forn11a,forn15a} that is in some sense a generalization of the traditional finite difference method, replacing the traditional polynomial basis functions with radial basis functions. The method is truly meshless, i.e.\ it can be used on arbitrarily distributed nodes.

This paper is organized as follows. In Section~\ref{sec:SDDForMaxwellsEquations} we present the mathematical background underlying the stochastic domain decomposition method for the two-dimensional Maxwell's equations. This includes both a discussion of the stochastic representation of the exact solution of Maxwell's equations and the description of the numerical evaluation of this representation in the context of stochastic domain decomposition. Section~\ref{sec:SDDImplementationMaxwellsEquations} details the numerical implementation of the stochastic domain decomposition method. This concerns both the choice for the discretization of the stochastic representation of the exact solution of Maxwell's equations, and the implementational details of the RBF-FD method. Numerical results for an analytical solution and some simple geophysical examples are presented in Section~\ref{sec:SDDForMaxwellsEquationsNumericalResults}. Although simple, these traditional tests show the potential of the stochastic domain decomposition method. The conclusion and final thoughts are given in Section~\ref{sec:ConlusionSDDMaxwellsEquations}.

\section{Stochastic domain decomposition for the\texorpdfstring{\\}{ }two-dimensional Maxwell's equations}\label{sec:SDDForMaxwellsEquations}

In this section we present the necessary theoretical background underlying the stochastic domain decomposition for the quasi-static two-dimensional Maxwell's equations as used in the magnetotelluric method.

\subsection{The two-dimensional Maxwell's equations}

The two-dimensional quasi-static Maxwell's in the time frequency domain read
\begin{align}\label{eq:MaxwellsEquation}
\begin{split}
 &\nabla\cdot\left(\frac1{i\omega\mu}\nabla E^y\right)-\sigma E^y=0,\qquad
 \nabla\cdot\left(\frac1{\sigma}\nabla H^y\right)-i\omega\mu H^y=0,
\end{split}
\end{align}
where $\nabla=(\partial_x,\partial_z)$ is the two-dimensional gradient operator in the $(x,z)$-plane, $E^y$ and $H^y$ are the $y$-components of the electric and magnetic field vectors $\mathbf{E}$ and $\mathbf{H}$, respectively, $\sigma$ is the electric conductivity, $\mu=\mu_0=4\cdot10^{-7}\, \textup{Hm}^{-1}$ is the magnetic permeability, $\omega$ is the angular frequency, and $i=\sqrt{-1}$ is the imaginary unit. The equation for the electric field component is called the TE-mode, whereas the equation for the magnetic field component is called the TM-mode.

\begin{figure}[!ht]
    \centering
    \includegraphics[width=0.4\textwidth]{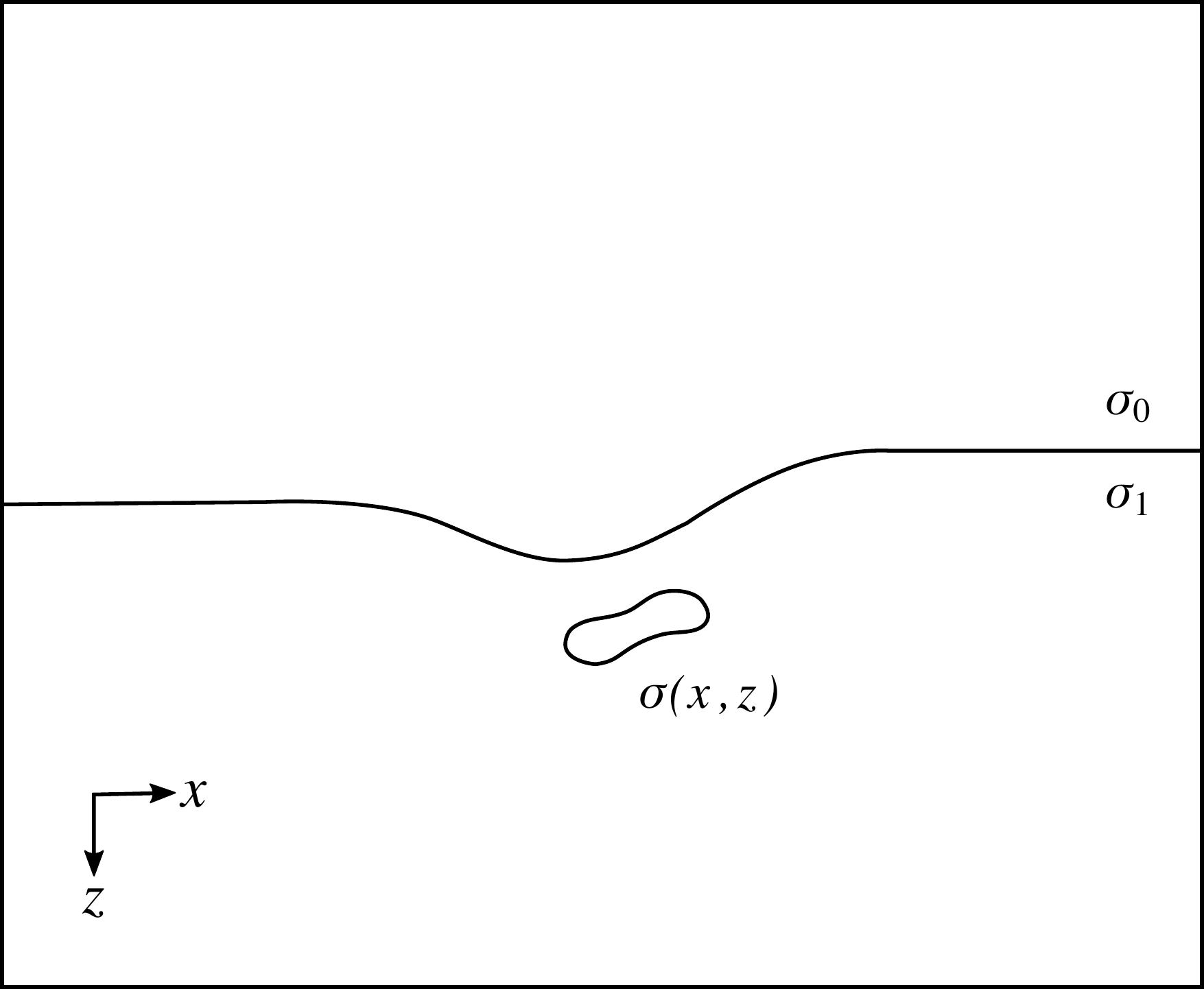}
    \caption{A sample subsurface structure and computational domain for system~\eqref{eq:MaxwellsEquation}.}
    \label{fig:plotOfDomain}
\end{figure}

The above system~\eqref{eq:MaxwellsEquation} gives the components of the primary fields perpendicular to the plane of the model, from which the secondary field components $E^x$ and $H^x$ in the plane of the model can be derived as
\begin{align}\label{eq:MaxwellsEquationSecondary}
\begin{split}
 &E^x=\frac1\sigma\pdl{H^y}{z},\qquad
 H^x=-\frac{1}{i\omega\mu}\pdl{E^y}{z}.
\end{split}
\end{align}
The system for the field components $E^x$, $E^y$, $H^x$ and $H^y$ has to be complemented with appropriate boundary conditions. In the following we will work with Dirichlet boundary conditions exclusively. More precisely, we will assume that all the boundaries are far away from any regions of anomalous conductivity so that the one-dimensional half-space boundary conditions can be used on the left and on the right of the domain~\cite[p.~56; Figure~\ref{fig:plotOfDomain}]{weav94a}. The top and bottom boundaries are obtained from linear interpolation from the respective top and bottom corner points of the domain, respectively.

The proposed stochastic domain decomposition method can also be applied to other kinds of boundary conditions, including Neumann and Robin boundary conditions. For further details, see~\cite{mair13a}.

Once the primary and secondary field components are computed, they can be used to calculate the apparent resistivities and phases as
\begin{align}\label{eq:ResistivityPhaseDefinition}
\begin{split}
 &\rho^{\rm TE}_a=\frac{1}{\omega\mu}\left|\frac{E^y}{H^x}\right|^2\quad\textup{and}\quad \varphi^{\rm TE}=\textup{arg}\left(\frac{E^y}{H^x}\right),\\
 &\rho^{\rm TM}_a=\frac{1}{\omega\mu}\left|\frac{E^x}{H^y}\right|^2\quad\textup{and}\quad \varphi^{\rm TM}=\textup{arg}\left(\frac{E^x}{H^y}\right).
\end{split}
\end{align}

\subsection{Stochastic analysis for the two-dimensional Maxwell's equations}\label{subsec:StochasticAnalysisMaxwellsEquations}

For a given domain $\Omega\subset\mathbb{R}^2$, it is well-known that for linear elliptic boundary value problems of the form
\begin{equation}\label{eq:GeneralEllipticEquation}
 \mathcal{L}u-\lambda(x,z) u=0\quad \textup{in}\ \Omega,\qquad u|_{\partial\Omega}=g(x,z),
\end{equation}
where $\lambda$ is non-negative and $\mathcal{L}=\frac12a_{ij}(x,z)\partial_i\partial_j + b_i(x,z)\partial_i$ is a semi-elliptic operator with bounded, two times continuously differentiable coefficients (summation over repeated indices is implied), the exact solution can be written in probabilistic form as
\begin{subequations}\label{eq:StochasticSolutionEllipticEquation}
\begin{equation}\label{eq:StochasticSolutionEllipticEquationA}
 u(x,z)=\mathrm{E}\left(g(\beta(\tau_{\partial\Omega}))\exp\left(-\int_0^{\tau_{\partial\Omega}}\lambda(\beta(s))\,\mathrm{d}s\right)\Big| \beta(0)=(x,z)\right).
\end{equation}
Here, $\beta(t)=(X(t),Z(t))$ denotes the stochastic process associated with the operator $\mathcal L$, satisfying the stochastic differential equation
\begin{equation}\label{eq:StochasticSolutionEllipticEquationB}
\mathrm{d}\beta=b(\beta)\mathrm{d}t+V(\beta)\mathrm{d}W,
\end{equation}
\end{subequations}
where the two-dimensional drift vector $b$ has components $b_1$, $b_2$, and the $2\times 2$ matrices $V=(V_{ij})$ and $a=(a_{ij})$ are related through $VV^{\rm T}=a$, and $W$ is two-dimensional Brownian motion. By $\tau_{\partial\Omega}$ we denote the first hitting time of the boundary of~$\Omega$ for a stochastic process $\beta(t)$ starting at point $(x,z)$. Eq.~\eqref{eq:StochasticSolutionEllipticEquationA} is the celebrated \emph{Kac--Feynman formula}~\cite{mile12Ay}.

The main problem in using the stochastic solution~\eqref{eq:StochasticSolutionEllipticEquation} is that the two-dimensional Maxwell's equations are not of the form of~\eqref{eq:GeneralEllipticEquation} since the conductivity~$\sigma$ is in general not a continuous function. It is therefore necessary to study the class of problems given by
\begin{equation}\label{eq:GeneralEllipticEquationDiscontinuous}
 \frac12\nabla\cdot(\kappa(x,z)\nabla u)-\lambda(x,z)u=0\quad \textup{in}\ \Omega,\qquad u|_{\partial\Omega}=g(x,z),
\end{equation}
where both $\kappa$ and $\lambda$ are discontinuous. For the two-dimensional Maxwell's equations we have $\kappa=2/(i\omega\mu)$ and $\lambda=\sigma$ for the TE-mode and $\kappa=2/\sigma$ and $\lambda=i\omega\mu$ for the TM-mode. That is, $\lambda$ is discontinuous for the TE-mode and $\kappa$ is discontinuous for the TM-mode, and both parameters can be complex-valued. Multiplying the equation for the electric mode with~$i$, it is sufficient to assume that $\kappa\in\mathbb{R}^+$ and $\lambda\in\mathbb{C}$.

The stochastic analysis of this class of problems is considerably more elaborate; available theoretical results seem to be mostly restricted to one-dimensional and real-valued cases. While it follows from applying a regularization argument, see e.g.~\cite{leja11a}, that the solution of Eq.~\eqref{eq:GeneralEllipticEquationDiscontinuous} is still given through the Kac--Feynman formula~\eqref{eq:StochasticSolutionEllipticEquationA}, finding a suitable stochastic process associated with the operator $\frac12\nabla\cdot(\kappa(x,z)\nabla)$ for general forms of discontinuous $\kappa$ appears to be an open problem.

On the other hand, the construction of numerical approximations to this problem for the case of $\kappa$ (and $\lambda$) being piecewise constant has been the subject of several investigations, especially for the case of $\lambda=0$. See~\cite{boss15a,leja13a,mair13a,tupp12a} for recent results. Different schemes have been proposed, which include so-called \emph{kinetic schemes}~\cite{leja10a}, \emph{mixing schemes}~\cite{leja13a}, schemes relying on \emph{occupation times}~\cite{leja11a} and schemes using ideas of \emph{finite differences}~\cite{leja13a,masc04a}. Here, we have chosen this last approach and will discuss it in more detail.

The main idea of all the above approaches is to split the domain into sub-domains~$\Omega_i$ over which both~$\kappa$ and~$\lambda$ are constant. On each sub-domain, Eq.~\eqref{eq:GeneralEllipticEquationDiscontinuous} reduces to
\begin{equation}\label{eq:HelmholtzEquationsub-domains}
 \frac12\kappa_i\Delta u-\lambda_iu=0\qquad \text {in } \Omega_i.
\end{equation}
This equation is of the form~\eqref{eq:GeneralEllipticEquation} and hence the stochastic solution as given by~\eqref{eq:StochasticSolutionEllipticEquation} holds. More precisely, the stochastic differential equation~\eqref{eq:StochasticSolutionEllipticEquationB} simplifies to
\begin{equation}\label{eq:StochasticSolutionEllipticEquationBSimplified}
 \mathrm{d}\beta=\sqrt{\kappa_i}\,\mathrm{I}_{\Omega_i}(\beta)\,\mathrm{d}W,
\end{equation}
where here and in the following~$\mathrm{I}_A$ is the indicator function of~$A$. This means that the process can be simulated via regular \textit{Brownian motion} when it is away from the interface. If it reaches the interface, it is possible that for a while the process
goes to and fro between adjacent sub-domains before it randomly
resolves in either direction. The issue is that when the diffusion coefficients are different, every time that the process crosses the interface, the regime changes. 

The approximation of the  $\beta(t)$ process as it passes through the interface can be based on finite differences by imposing the condition of continuity of the flux $\kappa\nabla u$ across the interface,
\begin{equation}\label{eq:ExactInterfaceCondition}
 \kappa_i\nabla u=\kappa_j\nabla u\quad \textup{at}\ \gamma_{ij},
\end{equation}
where $\gamma_{ij}$ is the interface between the neighboring sub-domains $\Omega_i$ and $\Omega_j$. 

Without loss of generality, we assume that there is only one vertical interface separating the two neighboring sub-domains~$\Omega_1$ and~$\Omega_2$ located at $x=0$. At this interface, the continuity of the flux condition~\eqref{eq:ExactInterfaceCondition} can be represented as
\begin{equation}
 \kappa_1\lim_{h_1\to0}\frac{u(h_1,z)-u(0,z)}{h_1}=
\kappa_2\lim_{h_2\to0}\frac{u(-h_2,z)-u(0,z)}{-h_2},\qquad \forall (0,z)\in\gamma,
\end{equation}
which can be locally solved for~$u(0,z)$ yielding
\begin{equation}\label{eq:StochasticFormulaAtInterface}
 u(0,z)\approx p_1u(h_1,z)+ p_2u(-h_2,z),\qquad h_1,\,h_2>0,
\end{equation}
with
\[
p_1=\frac{\kappa_1h_2}{\kappa_1h_2+\kappa_2h_1},\quad p_2=\frac{\kappa_2h_1}{\kappa_1h_2+\kappa_2h_1}.
\]
If~$h_1=h_2=h$, then,
\begin{equation}\label{eq:CrossingProbability}
  p_1=\frac{\kappa_1}{\kappa_1+\kappa_2},\quad p_2=\frac{\kappa_2}{\kappa_1+\kappa_2}.
\end{equation}
An interpretation of the above formula~\eqref{eq:StochasticFormulaAtInterface} is that when the $\beta(t)$ process  reaches the interface, locally, $u(0,z)$ can be approximated as the expected value of a random variable taking values $u(h_1,z)$ and $u(-h_2,z)$ with probabilities $p_1$ and $p_2$, respectively. This, in turn, also admits the following probabilistic interpretation: after reaching the interface, the process resumes from either, a point in the neighborhood of $(h_1,z)$ with probability $p_1$, or the neighborhood of $(-h_2,z)$ with probability $p_2$. This interpretation provides a straightforward algorithm to deal with the process when it hits the interface. Note that formula~\eqref{eq:StochasticFormulaAtInterface} is not exact as long as~$h_1,h_2$ are finite. 

An alternative to the previous procedure consists of
dissecting the behaviour of the process as it moves through the interface by using notions of \textit{Skew Brownian motion} and \textit{Brownian meander}, see~\cite{leja11a,leja12a} for further discussions. An algorithm using this alternative approach is computationally more complex and expensive than the one based on finite differences. However, our testing indicates that both algorithms produce qualitatively similar results. For this reason, we choose to present only the procedure which is easier to implement.

\section{Numerical implementation}\label{sec:SDDImplementationMaxwellsEquations}

In this section we present the details of the novel numerical implementation of stochastic domain decomposition for the two-dimensional Maxwell's equations in the frequency domain.

\subsection{Stochastic solver}

The pointwise stochastic solution procedure consists of simulating realizations of the path $\{\beta^{(n)}\}$,  $\beta^{(n)}=\big(\beta^{(n)}_x,\beta^{(n)}_z\big)$, a discretized version of the solution to the differential equation~\eqref{eq:StochasticSolutionEllipticEquationBSimplified} based on a modification of the \textit{Euler--Maruyama method}. 
The starting position, $\beta^0=(x,z)$, is the point at which the numerical solution is sought and the realization of the process is completed when the overall boundary of $\Omega$ is reached. Suppose that at the step $n$ the process is in the sub-domain $\Omega_i$. Then, we draw a provisional $\beta^{(n+1)}$ as
\begin{equation}\label{eq:DiscretizedProcess}
\beta^{(n+1)}=\beta^{(n)}+\sqrt{\kappa_i\,\Delta t}\, W, \qquad W\sim\mathcal{N}(\mathbf{0},I_2)
\end{equation}
where $\Delta t=t^{(n+1)}-t^{(n)}$ is the time step, and $\mathcal N(\mathbf{0},I_2)$ denotes the distribution of a random $2$-vector of independent standard normal variables. Next, we check if $\partial\Omega$ has been reached. If so, the realization is completed. Otherwise, we verify whether  between the times $t^{(n)}$ and $t^{(n+1)}$ the interface has been hit.  If the process is away from the interface, $\beta^{(n+1)}$ is retained and $\Delta t$ is added to $T_i$, the occupation time of $\Omega_i$, before moving on to the next iteration.

When the process hits an interface point $(x,z)|_{\gamma_{ij}}$, say $(x_{\gamma_{ij}},z)$, if $\kappa_i\neq\kappa_{j}$, we carry out the procedure laid out in Section~\ref{subsec:StochasticAnalysisMaxwellsEquations}, i.e., by using the probabilities $p_i$ and $p_{j}=1-p_i$ from~\eqref{eq:CrossingProbability}, we randomize to determine whether $\beta^{(n+1)}$ lies  in $\Omega_i$ or in $\Omega_{j}$. Also, we estimate the occupation times between $t^{(n)}$ and $t^{(n+1)}$. 
First, by either inverting the test for the first hitting time during an excursion with respect to a Brownian bridge (i.e., when the provisional $\beta^{(n+1)}$ is on the same sub-domain as $\beta^{(n)}$), or approximating the expected value of the first exit time with respect to a Brownian bridge, we estimate the time when the interface was first reached~\cite{buch05a,gobe00a}. We define $t^{(n)}_{\gamma_{ij}}$ as
\begin{equation}
t^{(n)}_{\gamma_{ij}}=\begin{cases}
  \dfrac{-2(\beta^{(n)}_x - x_{\gamma_{ij}})(\beta^{(n+1)}_x - x_{\gamma_{ij}})
  }{\kappa_i\,\log U_{\gamma}},& \quad U_{\gamma}\sim\mathcal{U}(0,1), \ \text{ if }  \beta^{(n)},\beta^{(n+1)} \text{ in } \Omega_i\\[12pt]
  \Delta t\,\left|\dfrac{\beta^{(n)}_x - x_{\gamma_{ij}}}{\beta^{(n+1)}_x-\beta^{(n)}_x}\right|, 
&\quad  \text{ if } \beta^{(n)} \text{ in } \Omega_i \text{ and } \beta^{(n+1)} \text{ in } \Omega_j,\label{eq:firstexittime}
\end{cases}
\end{equation}
where $\mathcal U(0,1)$ denotes the uniform distribution on $[0,1]$, so that whenever $t^{(n)}_{\gamma_{ij}}\leq\Delta t $, we conclude that
the process has reached the interface at time $t^{(n)}+t^{(n)}_{\gamma_{ij}}$ and, accordingly, proceed to randomize in order to
find the definitive value for $\beta^{(n+1)}$. 
This randomization and the update of the occupation times can be done as follows. Let
$c_{ij}(h)$ be
\[
c_{ij}(h)=h\,\mathrm{I}_{\left[\beta^{(n)}_x \leq
    x_{\gamma_{ij}}\right]} - h\,\mathrm{I}_{\left[\beta^{(n)}_x >
    x_{\gamma_{ij}}\right]}.
\]
Now, if $U_\alpha<p_i$, $U_\alpha\sim\mathcal{U}(0,1)$, then
$\beta^{(n+1)}$ is drawn randomly from inside a circle of radius $h_i$
and center $\left(x_{\gamma_{ij}}-c_{ij}(h_i),z\right)$, and
$\Delta t - \frac{1}{2}\left(\Delta t -
  t^{(n)}_{\gamma_{ij}}\right)p_i$ and
$\frac{1}{2}\left(\Delta t - t^{(n)}_{\gamma_{ij}}\right)p_i$ are
added to $T_i$ and $T_j$, respectively. Otherwise, $\beta^{(n+1)}$ is
drawn from a circle with radius $h_j$ and center
$\left(x_{\gamma_{ij}}+c_{ij}(h_j),z\right)$, and
$t^{(n)}_{\gamma_{ij}}+ \frac{1}{2}\left(\Delta t -
  t^{(n)}_{\gamma_{ij}}\right)p_j$ and
$\frac{1}{2}\left(\Delta t - t^{(n)}_{\gamma_{ij}}\right)(1+p_j)$ are
added to $T_i$ and $T_j$, respectively. These occupation times
estimates are based on approximations of their expected value as the
process moves around the interface given $\gamma_{ij}$~\cite{leja11a}.  When $\kappa_i=\kappa_j$ and the
process hits the $\gamma_{ij}$ interface, there is no regime change
and we retain $\beta^{(n+1)}$ obtained through
\eqref{eq:DiscretizedProcess}. Only the occupation times between
$t^{(n)}$ and $t^{(n+1)}$ are updated according to where~$\beta^{(n+1)}$ lies.

This process is repeated $N$ times so that the expected value appearing in~\eqref{eq:StochasticSolutionEllipticEquationA} can be approximated as
\[
 u(x,z)=\frac1N\sum_{r=1}^Ng(\beta_r(\tilde \tau_r))\exp\left(-\sum_{j=1}^K\lambda_k T_j^r\right),
\]
where $\beta_r(\tilde\tau_r)$ denotes the state of the process at the
time $\tilde\tau_r$ of the $r$th realization of the discretized
process~\eqref{eq:DiscretizedProcess}, $\tilde \tau_r$ is the time
estimate at which the overall boundary, $\partial\Omega$, was first
reached, $K$ is the number of sub-domains $\Omega_j$ with different
$\kappa_j$ (and/or $\lambda_j$) and $T^r_j$ is the time spent in the
$j$th sub-domain through the $r$th realization. When $\partial\Omega$ consists of vertical or horizontal barriers, the estimation of
$\tilde \tau_r$, the first exit time from $\Omega$ for 
the $r$th realization of the process, can be carried through
an expression similar to \eqref{eq:firstexittime} with
$(x,z)|_{\gamma_{ij}}$ replaced by $(x,z)|_{\partial\Omega}$. Thus, when
$t^{(n)}_{\partial\Omega}\leq \Delta t$, we
conclude that the process has reached the boundary at time
$\tilde \tau_r=t^{(n)}+t^{(n)}_{\partial\Omega}$ bringing the $r$th realization of the process to an end.

\begin{remark}
 While the main aim of this paper is to put forward a new method for solving the two-dimensional Maxwell's equations over the entire computational domain, the stochastic form of the solution of Maxwell's equations can also be used to compute the solution in single, isolated points. Thus, if the solution is only required near the measurement sites, the stochastic solution can be used for this purpose as well. This is in striking contrast to all deterministic methods, which require the computations to be carried out over the entire domain, even if the solution is only required in a single point. An example for this use will be presented in Section~\ref{sec:SDDForMaxwellsEquationsNumericalResults}.
\end{remark}

\subsection{Sub-domain solver}

Since we are primarily interested in the application of the stochastic domain decomposition methods to the two-dimensional Maxwell's equations for realistic sub-surface models, the single sub-domains $\Omega_i$ will generally be of an arbitrary shape. This makes it natural to choose a sub-domain solver that can operate on an arbitrary set of mesh points distributed in a suitable way over the single sub-domain. This makes \emph{meshless methods} a suitable choice.

The underlying paradigm of meshless methods is that one does not require a topologically connected mesh in order to compute a numerical solution over a given domain, as is required in conventional mesh-based methods such as finite differences, finite volumes or finite elements. In meshless methods, the local approximations of derivatives are formulated directly in terms of (in principle) arbitrarily distributed nodes. Various meshfree methods have been developed over the past few decades, including smooth particle hydrodynamics, meshless finite differences, radial basis function methods, meshless local Petrov--Galerkin methods and element-free Galerkin methods~\cite{ding04Ay,forn15a,mile12Ay,nguy08Ay}.

In this paper we will use the radial basis function based finite differences method (RBF-FD). This method rests on replacing the one-dimensional polynomial test functions that are used to derive regular finite difference formulas by radial basis functions (RBFs)~\cite{forn11a}. More precisely, the weights $w_i$ at the node locations $x_i$, $i=1,\dots,n$ required to approximate a linear differential operator $\mathcal{L}$ at the node $x_0$ are obtained by solving the matrix system
\[
 \left(\begin{array}{cccc}\phi(||x_1-x_1||) & \phi(||x_1-x_2||) & \cdots & \phi(||x_1-x_n||)\\
 \phi(||x_2-x_1||) & \phi(||x_2-x_2||) & \cdots & \phi(||x_2-x_n||)\\
 \vdots & \vdots & & \vdots\\
 \phi(||x_n-x_1||) & \phi(||x_n-x_2||) & \cdots & \phi(||x_n-x_n||)\\
 \end{array}\right)\left(\begin{array}{c}w_1\\ w_2\\ \vdots\\ w_n \end{array}\right)=\left(\begin{array}{c}\mathcal{L}\phi(||x-x_1||)|_{x=x_0}\\ \mathcal{L}\phi(||x-x_2||)|_{x=x_0}\\ \vdots\\ \mathcal{L}\phi(||x-x_n||)|_{x=x_0} \end{array}\right).
\]
In other words, the weights are found in the RBF-FD method by requiring that the approximation of $\mathcal L$ is exact when applied to the radial basis functions themselves. Here, $x_i=(x_i^1,x_i^2,\cdots,x_i^d)^{\rm T}$ is a point in $d$-dimensional space, $||\cdot||$ is the Euclidean 2-norm and $\phi(||x-x_i||)$ is an RBF centered at the node $x_i$.

Common RBFs used in practice include the Gaussian, $\phi(r)=\exp(-(\ve r)^2)$, the multiquadric, $\phi(r)=\sqrt{1+(\ve r)^2}$, and the inverse multiquadric, $\phi(r)=(1+(\ve r)^2)^{-1/2}$~\cite{forn11a}. The parameter $\ve$ is called the shape parameter and it controls the flatness of the RBF. In the following, we will use the multiquadric for all computations.

For the given RBFs the above matrix system can be solved provided that the nodes $x_i$, $i=1,\dots,n$ are distinct. Denoting by $A$ the matrix with elements $A_{ij}=\phi(||x_i-x_j||)$ and by $B$ the square matrix with elements $B_{ij}=\mathcal{L}\phi(||x-x_j||)|_{x=x_i}$ the differentiation matrix~$\mathcal D$ approximating the linear operator $\mathcal L$ becomes
\[
 \mathcal{D}=B^{\rm T}A^{-1}.
\]
In other words, for the action of~$\mathcal L$ on a function~$f(x)$, we have the following approximation
\[
 \left(\begin{array}{c}\mathcal{L}f(x)|_{x=x_1}\\ \mathcal{L}f(x)|_{x=x_2}\\ \vdots\\ \mathcal{L}f(x)|_{x=x_n}\end{array}\right)\approx \mathcal{D}\left(\begin{array}{c}f(x_1)\\ f(x_2)\\ \vdots\\ f(x_n)\end{array}\right).
\]

A main problem with the above procedure is that the differentiation matrix $\mathcal D$ is a full matrix. Its computation requires $O(n^3)$ operations, which is very costly if~$\mathcal D$ has to be re-computed. This happens, for example, if the node layout $\{x_i\}$ changes, which is always necessary for moving mesh methods or if adaptive mesh refinement is used.

A more cost efficient way is achieved by assigning to each of the~$n$ nodes, $x_i$, a separate stencil of $n_s\ll n$ nodes. These nodes are typically the $n_s-1$ nearest neighbors of each node $x_0$. In this procedure, the differentiation matrix $\mathcal D$ becomes a sparse matrix having $n_s$ non-zero entries in each of the $n$ rows. This restriction to neighboring nodes yields the RBF-FD method. It is the exact analogue of the classical finite difference method, using RBFs instead of polynomials as basis functions in the stencil around each point $x_0$.

For Maxwell's equations, we require to approximate $\mathcal L=\Delta$ on each sub-domain with continuous conductivity. We do this by creating separate differentiation matrices for $\mathcal L=\partial_x^2$ and $\mathcal L=\partial_z^2$ using $n_s=9$ nodes per each stencil. These nodes are chosen to be the $8$ nearest neighbors and the center node $x_0$ itself.

The choice of the shape parameter~$\ve$ is paramount in that it governs the accuracy of the RBF-FD method. It is generally found that the numerical computations become most accurate when using almost flat RBFs (i.e.\ $\ve$ being very small). The smaller the parameter~$\ve$, however, the more ill-conditioned the matrix systems become~\cite{forn13a,wrig06Ay}. In order to overcome the dilemma of choosing between accuracy and ill-conditioning, several methods have been proposed, including the use of high precision arithmetics and the RBF-QR method~\cite{forn11b}, which was mostly developed for Gaussian RBFs. We found experimentally that a value of $\ve=1/2000$ gives satisfying accuracy for a wide range of grid spacings in the test problems considered in the following section. A more thorough investigation of the optimal shape parameter~$\ve$ for use within the stochastic domain decomposition method for Maxwell's equations should be investigated elsewhere.

\section{Results}\label{sec:SDDForMaxwellsEquationsNumericalResults}

In this section we present numerical results using the stochastic solution technique for the two-dimensional Maxwell's equations discussed in the previous section. These examples are well-studied in the literature and serve as a demonstration for the potential of the new method to correctly reproduce existing results. More realistic sub-surface models that will also demonstrate of the full potential of the meshless RBF-FD method will be presented in a separate paper.

\subsection{Analytical test model}

\begin{figure}[!ht]
    \centering
    \begin{subfigure}[b]{0.5\textwidth}
        \centering
        \includegraphics[width=\textwidth]{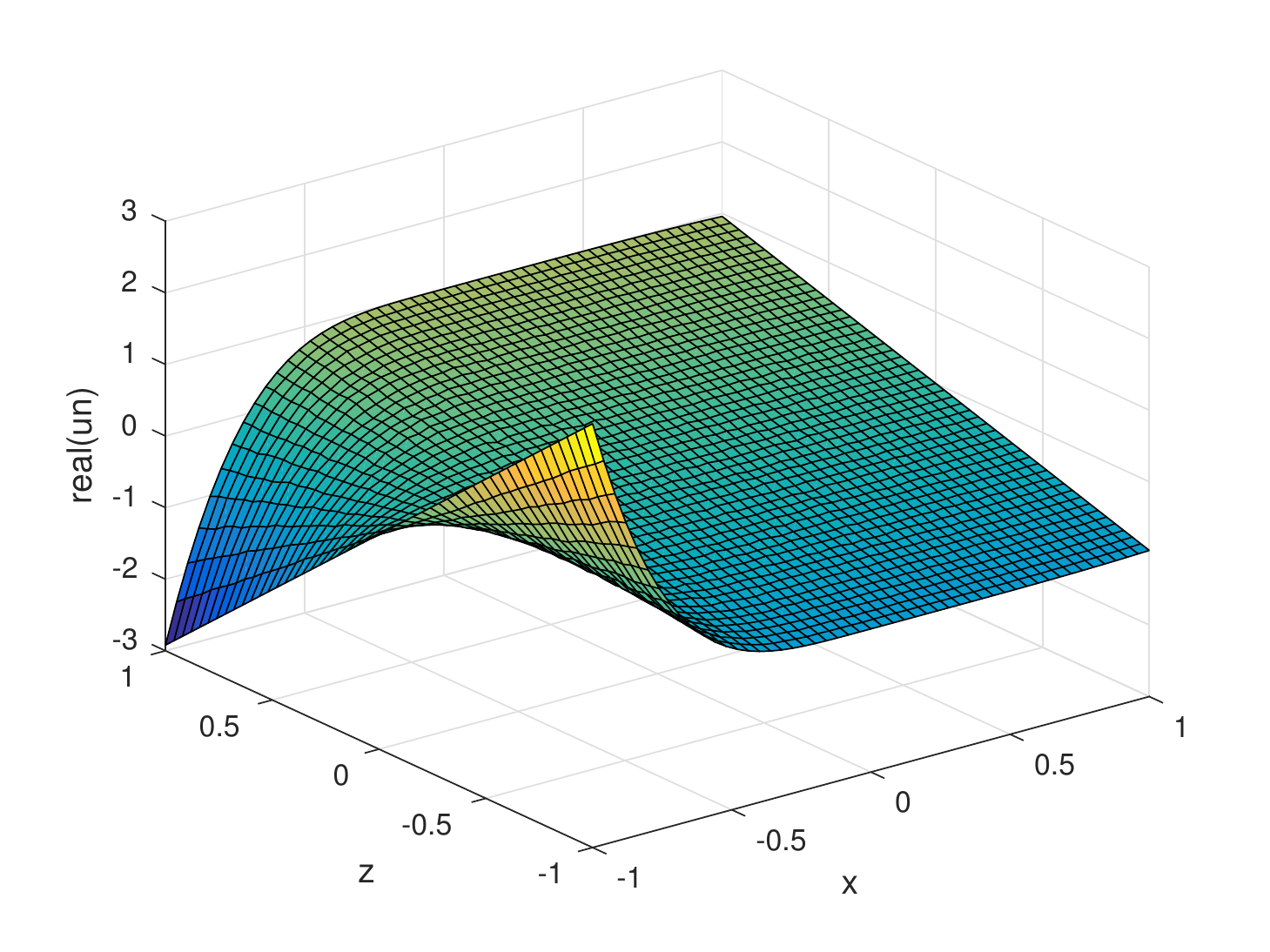}
    \end{subfigure}~
    \begin{subfigure}[b]{0.5\textwidth}
        \centering
        \includegraphics[width=\textwidth]{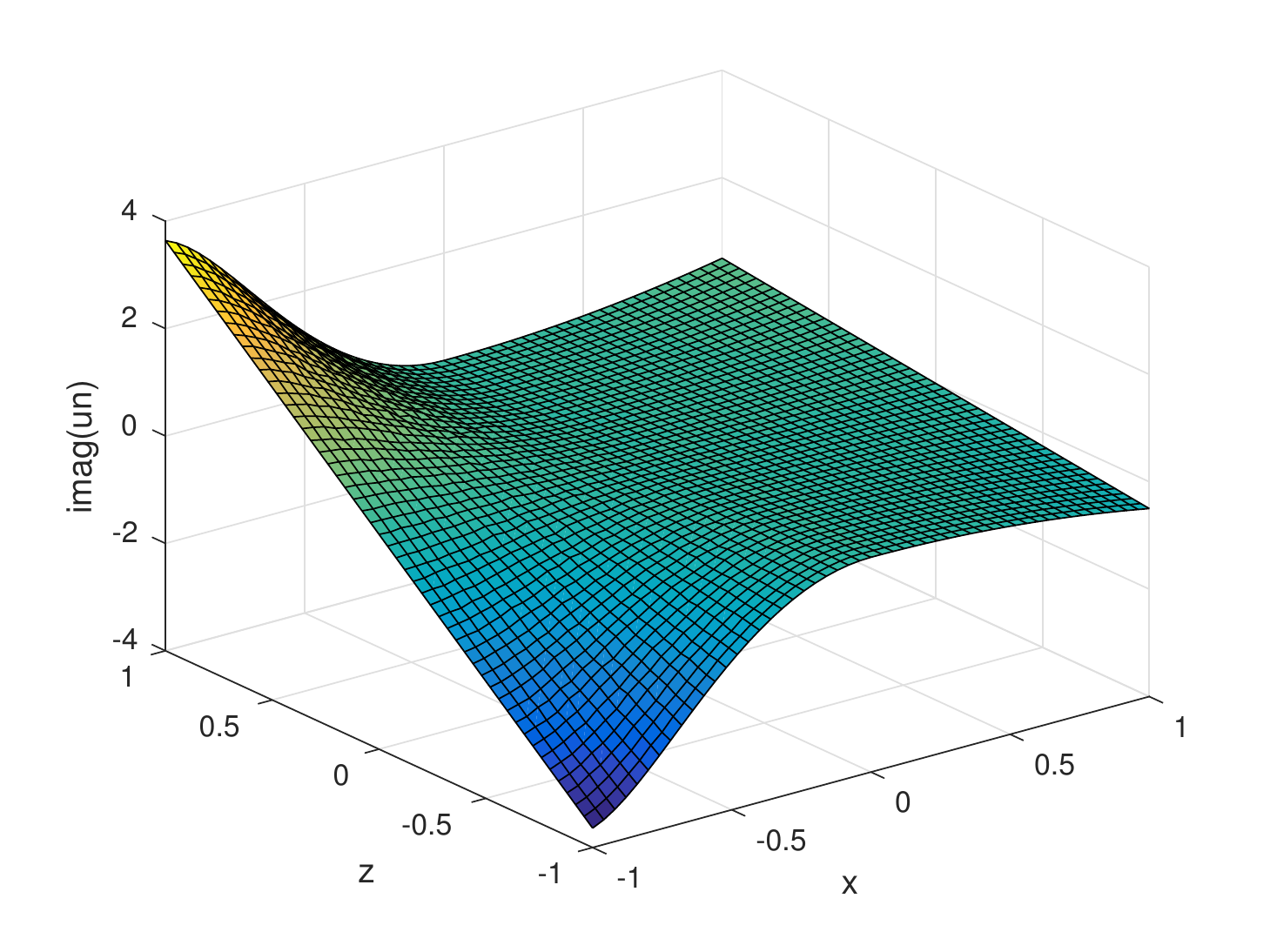}
    \end{subfigure}
    \caption{Numerical solution for the analytical test problem using $N=10000$ Monte-Carlo simulations. Left: Real part, right: Imaginary part.}
    \label{fig:AnalyticalTestModel}
\end{figure}

\begin{figure}[!ht]
    \centering
    \begin{subfigure}[b]{0.5\textwidth}
        \centering
        \includegraphics[width=\textwidth]{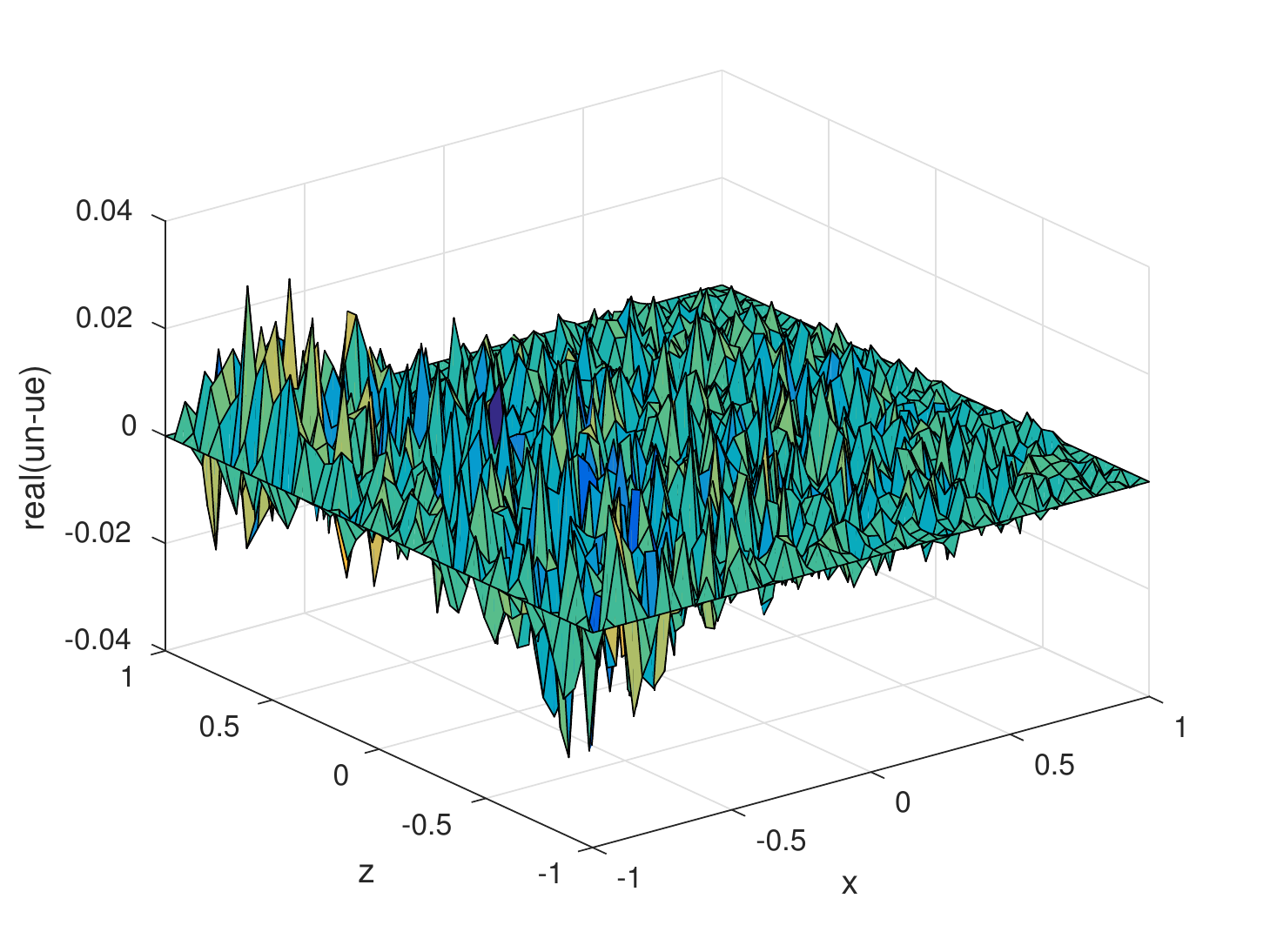}
    \end{subfigure}~
    \begin{subfigure}[b]{0.5\textwidth}
        \centering
        \includegraphics[width=\textwidth]{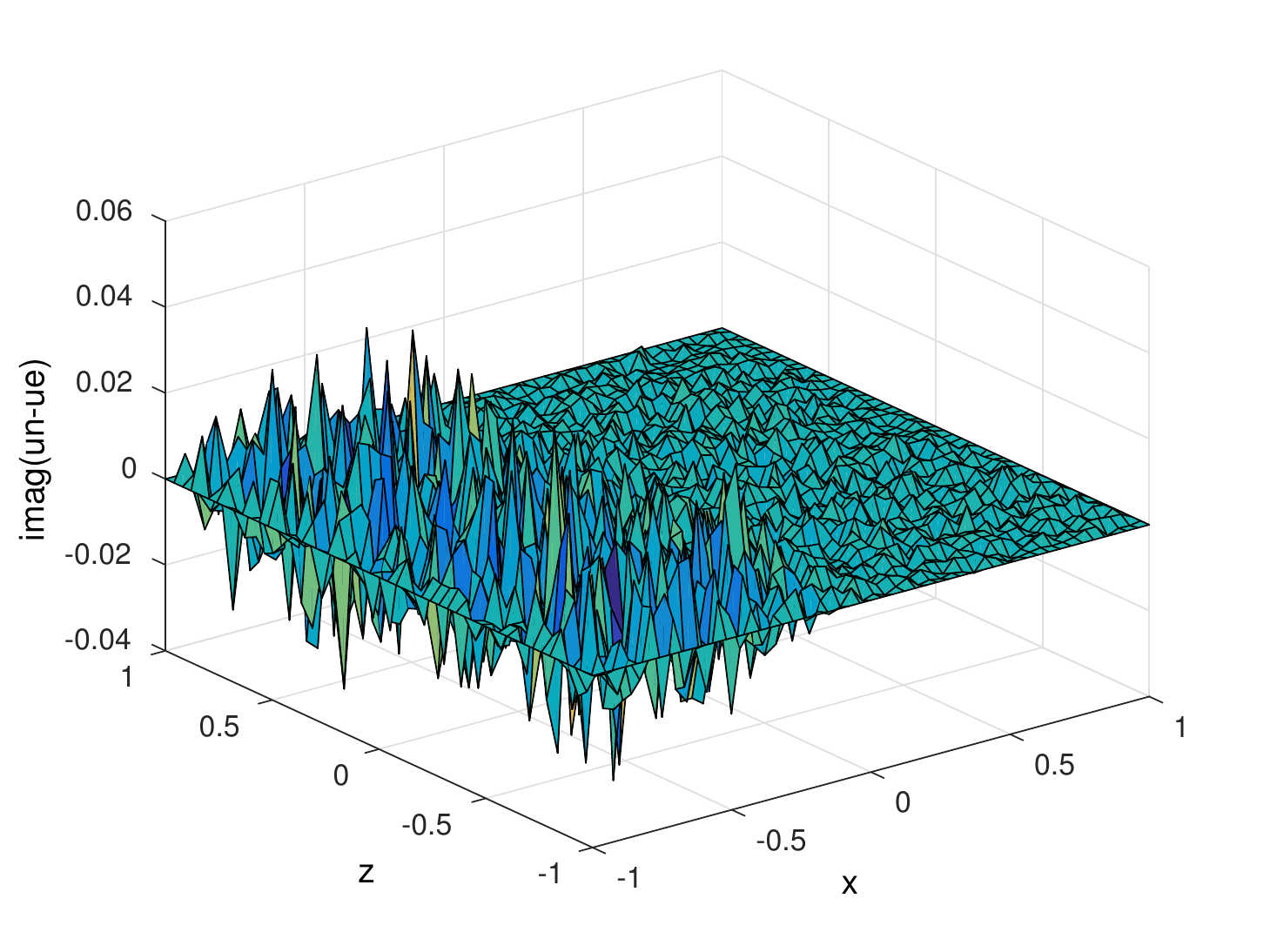}
    \end{subfigure}
    \caption{Error plots for the analytical test problem for $N=10000$ Monte-Carlo simulations. Left: Real part, right: Imaginary part.}
    \label{fig:AnalyticalTestModelError}
\end{figure}

To verify numerically our method for evaluating the process as it passes through the interface for discontinuous~$\kappa$ and imaginary~$\lambda$, we consider the Dirichlet problem for the complex-valued Helmholtz equation
\begin{subequations}\label{eq:analyticalTestModel}
\begin{equation}
 \nabla\cdot (\kappa \nabla)u+\lambda u=0,
\end{equation}
with exact solution
\begin{equation}
 u_{\rm e}= c_1(z+c_2)\left\{\begin{array}{cc} \cosh\left(\sqrt{\frac{\lambda}{\kappa_1}x}\right), & -1\le x< 0\\ \cosh\left(\sqrt{\frac{\lambda}{\kappa_2}x}\right), & 0\le x\le 1. \end{array}\right.
\end{equation}
\end{subequations}
Here we choose $c_1=c_2=1$, $\kappa_1=1$, $\kappa_2=10$ and $\lambda=10i$ as parameters. The exact solution~$u_{\rm e}$ is used as boundary data for the physical domain. The physical domain $\Omega=[-1,1]\times [-1,1]$ is discretized using a uniform mesh with $51\times 51$ grid points. The stochastic differential equation~\eqref{eq:StochasticSolutionEllipticEquationB} is discretized using the Euler--Maruyama method with time step~$\Delta t\propto(\Delta x)^2$. In Fig.~\ref{fig:AnalyticalTestModel} we display the numerical solution~$u_{\rm n}$ obtained from using the stochastic procedure in all grid points with $N=10 000$ Monte-Carlo simulations. The associated point-wise errors are displayed in Fig.~\ref{fig:AnalyticalTestModelError}.

In order to show the decrease of the errors as the number of Monte-Carlo simulations, $N$, is increased, we present the absolute errors for different values of~$N$ in Table~\ref{tab:AnalyticalTestModelResults}. Since numerically solving~\eqref{eq:analyticalTestModel} at every point in~$\Omega$ is computationally intensive for a study of increasing~$N$, the errors reported in Table~\ref{tab:AnalyticalTestModelResults} are for the single point $(x,z)=(0.6,0.6)$ only. While the magnitude of the error is spatially dependent, the error decrease with increasing~$N$ demonstrated below at $(0.6,0.6)$ remains valid throughout the domain.

\begin{table}[!ht]
\centering
\caption{Absolute errors for the model~\eqref{eq:analyticalTestModel} at point $(0.6,0.6)$ varying~$N$.}
\begin{tabular}{cccc}
\hline
$N$ & $1\cdot 10^{4}$ & $1\cdot 10^{5}$ & $1\cdot 10^{6}$\\
\hline
 $\Re (\textup{error})$ & 0.0086 & 0.0017 & $7.25\cdot 10^{-4}$ \\
 $\Im (\textup{error})$ & 0.0067 & 0.0034 & $8.95\cdot 10^{-4}$ \\
\hline
\end{tabular}
\label{tab:AnalyticalTestModelResults}
\end{table}

The convergence results presented in Table~\ref{tab:AnalyticalTestModelResults} should be taken with a grain of salt. Recall that the error incurred by numerically evaluating the stochastic representation of an exact solution of a linear boundary value problem such as~\eqref{eq:StochasticSolutionEllipticEquation} consists of three parts. These are the pure \emph{Monte-Carlo error} (due to approximating the expected value in Eq.~\eqref{eq:StochasticSolutionEllipticEquationA} with the mean value), the \emph{time stepping error} (due to discretizing the stochastic differential equation~\eqref{eq:StochasticSolutionEllipticEquationB} using a finite time step), and the \textit{error in estimating the first exit time}~$\tau_{\partial\Omega}$~\cite{aceb05a}. Increasing only the number of Monte-Carlo simulations as done in Table~\ref{tab:AnalyticalTestModelResults} hence will not lead to a convergent numerical scheme unless also the two other sources of errors are controlled, e.g.\ by using increasingly small time steps which will both reduce the time stepping error and improve the estimate for the first exit time. What Table~\ref{tab:AnalyticalTestModelResults} does demonstrate is that if a reasonably small time step is chosen in the discretization of the stochastic differential equation~\eqref{eq:StochasticSolutionEllipticEquationB} (controlling the second and third sources of numerical error), the numerical results obtained can be improved by merely increasing the number of Monte-Carlo simulations. This also shows that the additional error introduced due to our approximation strategy at the interface is small enough to prevent error saturation before geophysically acceptable accuracy is achieved. This is also explicitly demonstrated in the following examples.

\subsection{Quarter-space solution}

\begin{figure}[!ht]
  \centering
  \includegraphics[width=0.5\textwidth]{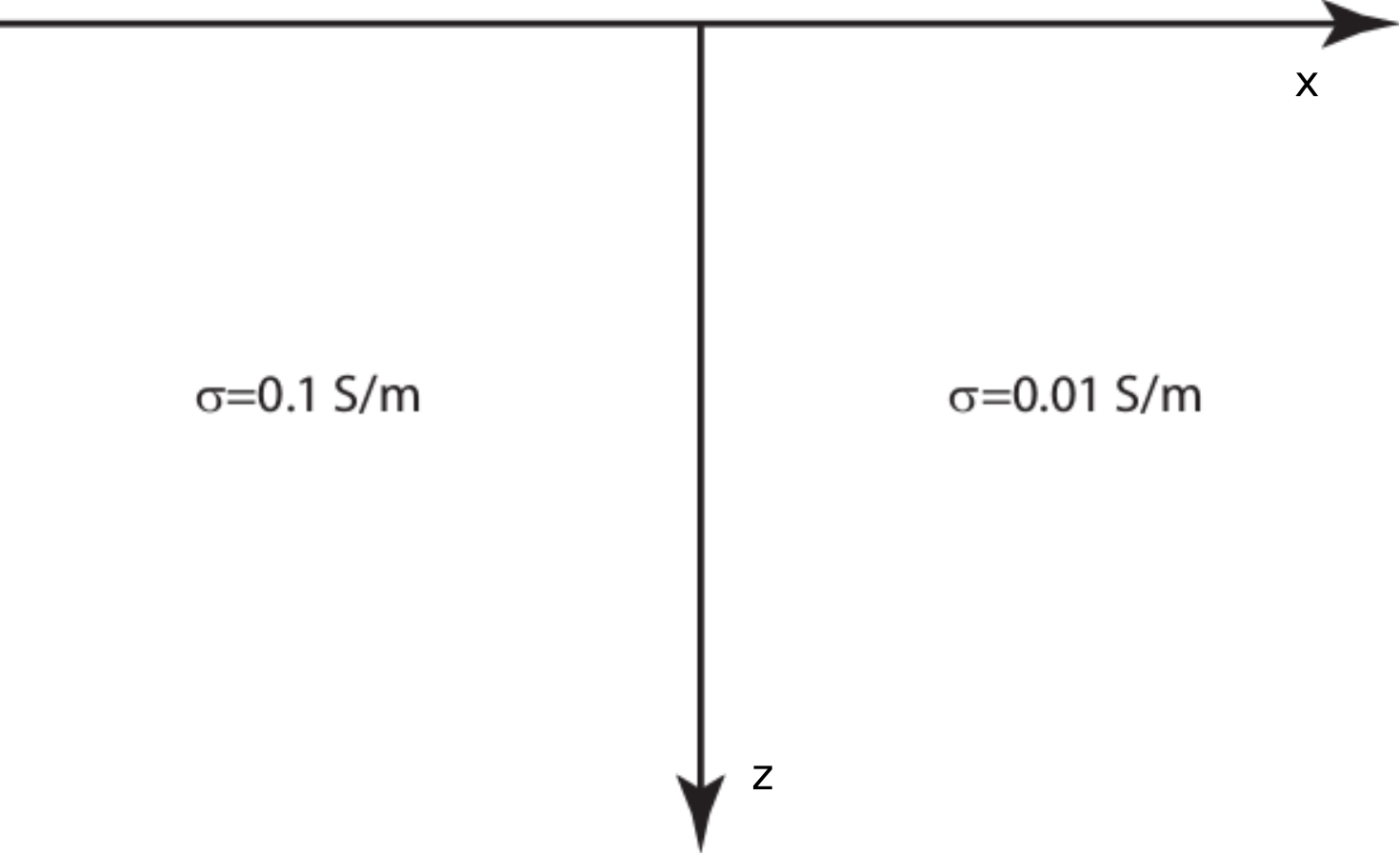}\\
  \caption{Conductivity model for the quarter-space experiment.}
  \label{fig:QuarterSpaceConductivityModel}
\end{figure}

The quarter-space model for this experiment is identical to the one proposed in~\cite{fisc92a,fisc93a}. It splits the region $z\ge0$ into two areas, one with conductivity $\sigma=0.1\, S/\textup{m}$ (on the left), the other with conductivity $\sigma=0.01\, S/\textup{m}$ (on the right), see Fig.~\ref{fig:QuarterSpaceConductivityModel}.

As in~\cite{fisc92a,fisc93a}, we used $f=1\,\textup{Hz}$ as the frequency. We employed a variable grid spacing with minimum cell sizes being $\Delta x\times \Delta z=50\,\textup{m}\times 50\,\textup{m}$ near the interfaces and a maximum cell size of $\Delta x\times \Delta z=300\,\textup{m}\times 200\,\textup{m}$ near the boundaries in the ground. A total of $N=5000$ Monte-Carlo simulations was used in the stochastic solver, which here and in the following was only used at the sub-domain interfaces, with the solution over the sub-domains being computed using the deterministic, meshless solver. Note that the variable resolution of the model is naturally handled using the meshless solver.

\begin{figure}[!ht]
    \centering
    \begin{subfigure}[b]{0.5\textwidth}
        \centering
        \includegraphics[width=\textwidth]{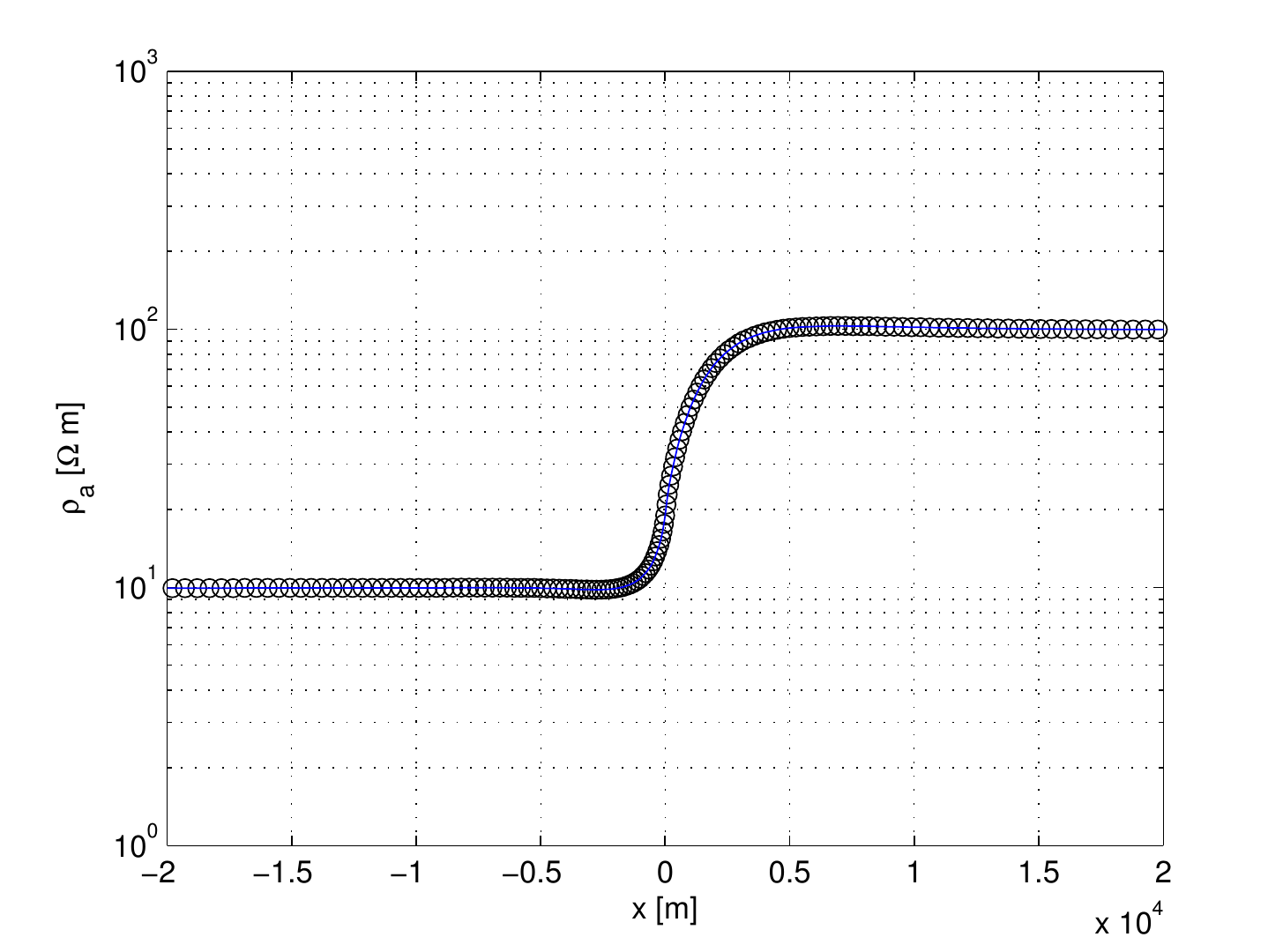}
    \end{subfigure}~
    \begin{subfigure}[b]{0.5\textwidth}
        \centering
        \includegraphics[width=\textwidth]{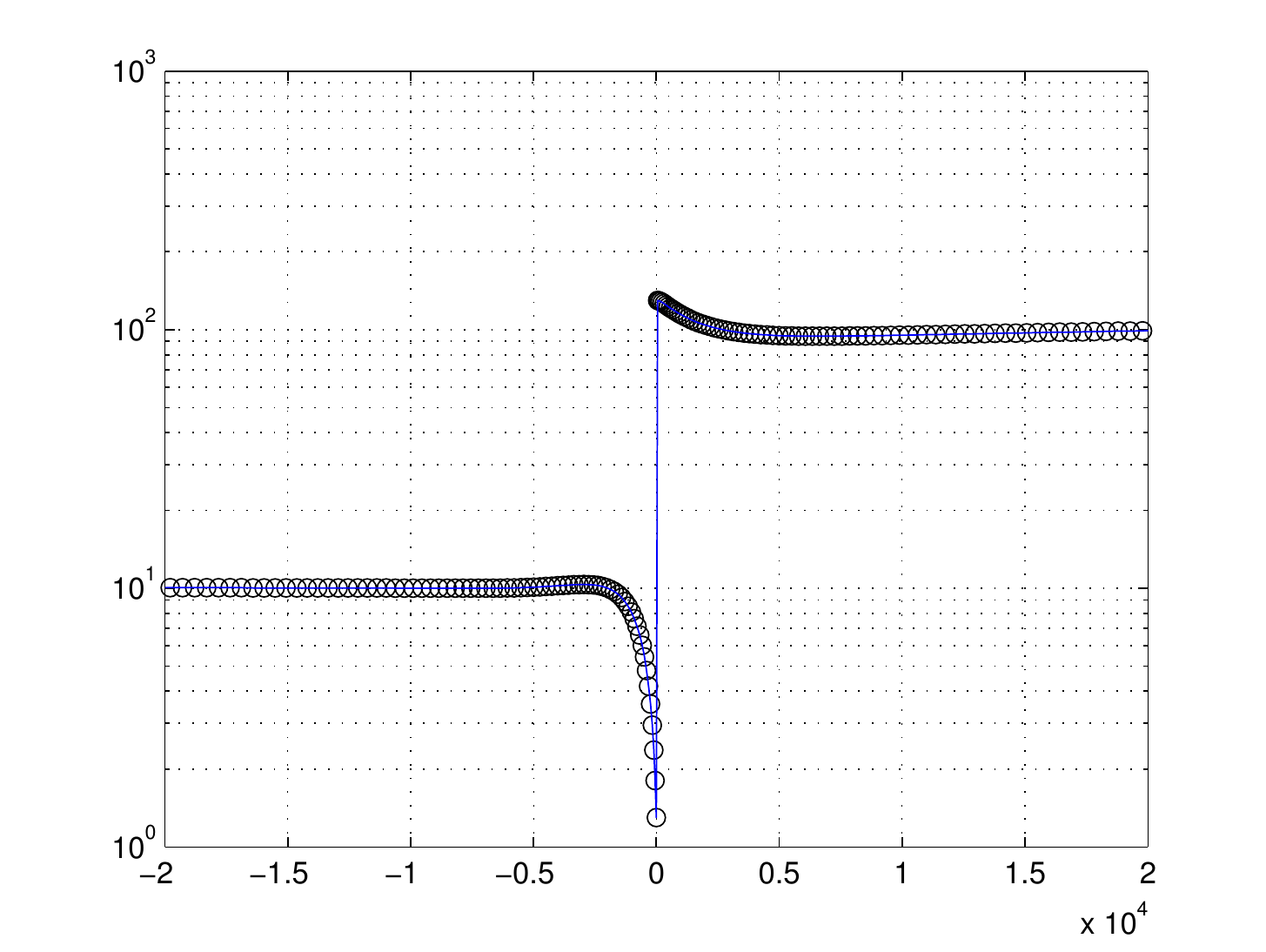}
    \end{subfigure}
    \caption{Apparent resistivities for the TE-mode (left) and the TM-mode (right) for the quarter-space model using $f=1\,\textup{Hz}$.}
    \label{fig:QuarterSpaceResistivities}
\end{figure}

\begin{figure}[!ht]
    \centering
    \begin{subfigure}[b]{0.5\textwidth}
        \centering
        \includegraphics[width=\textwidth]{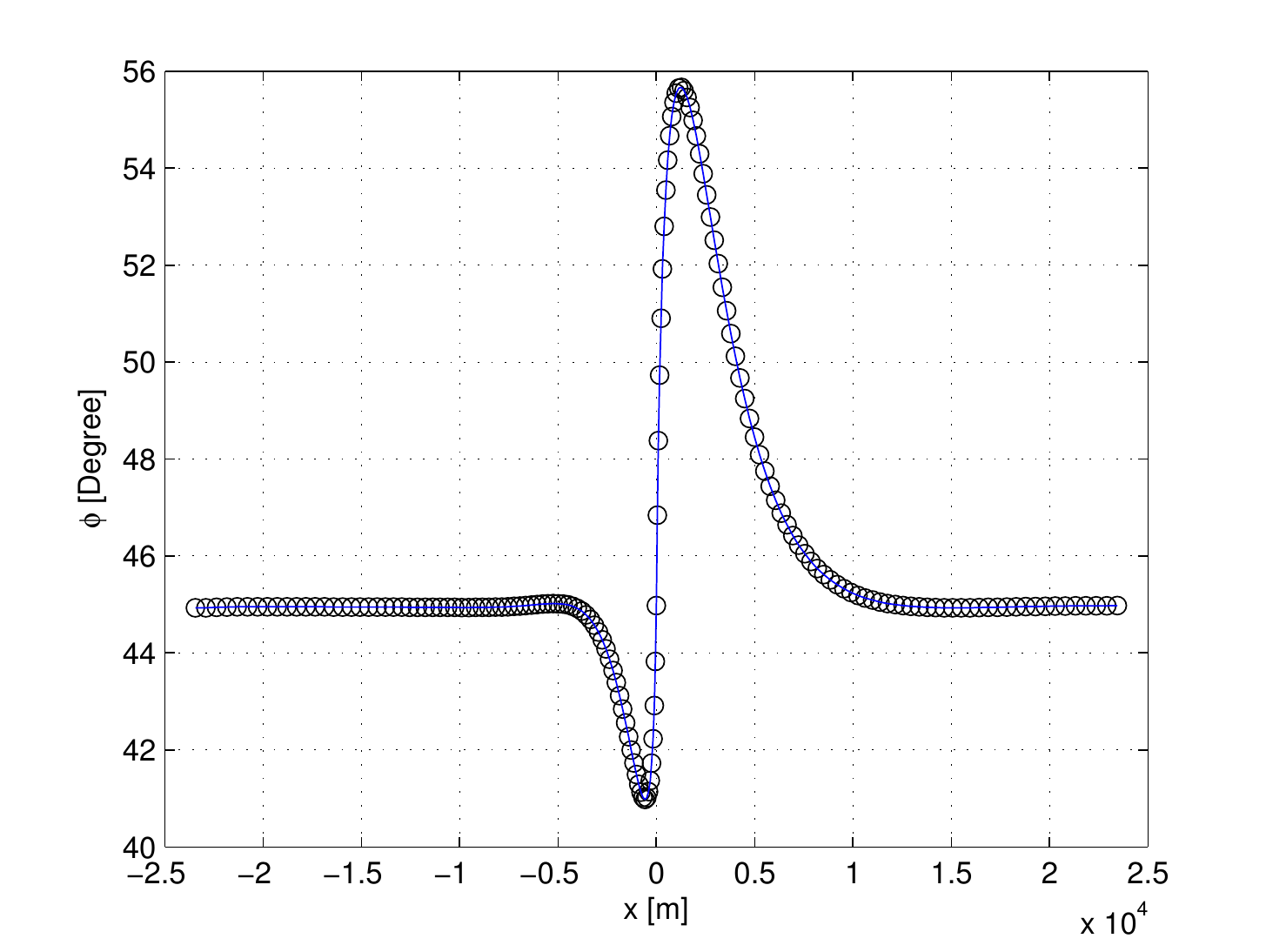}
    \end{subfigure}~
    \begin{subfigure}[b]{0.5\textwidth}
        \centering
        \includegraphics[width=\textwidth]{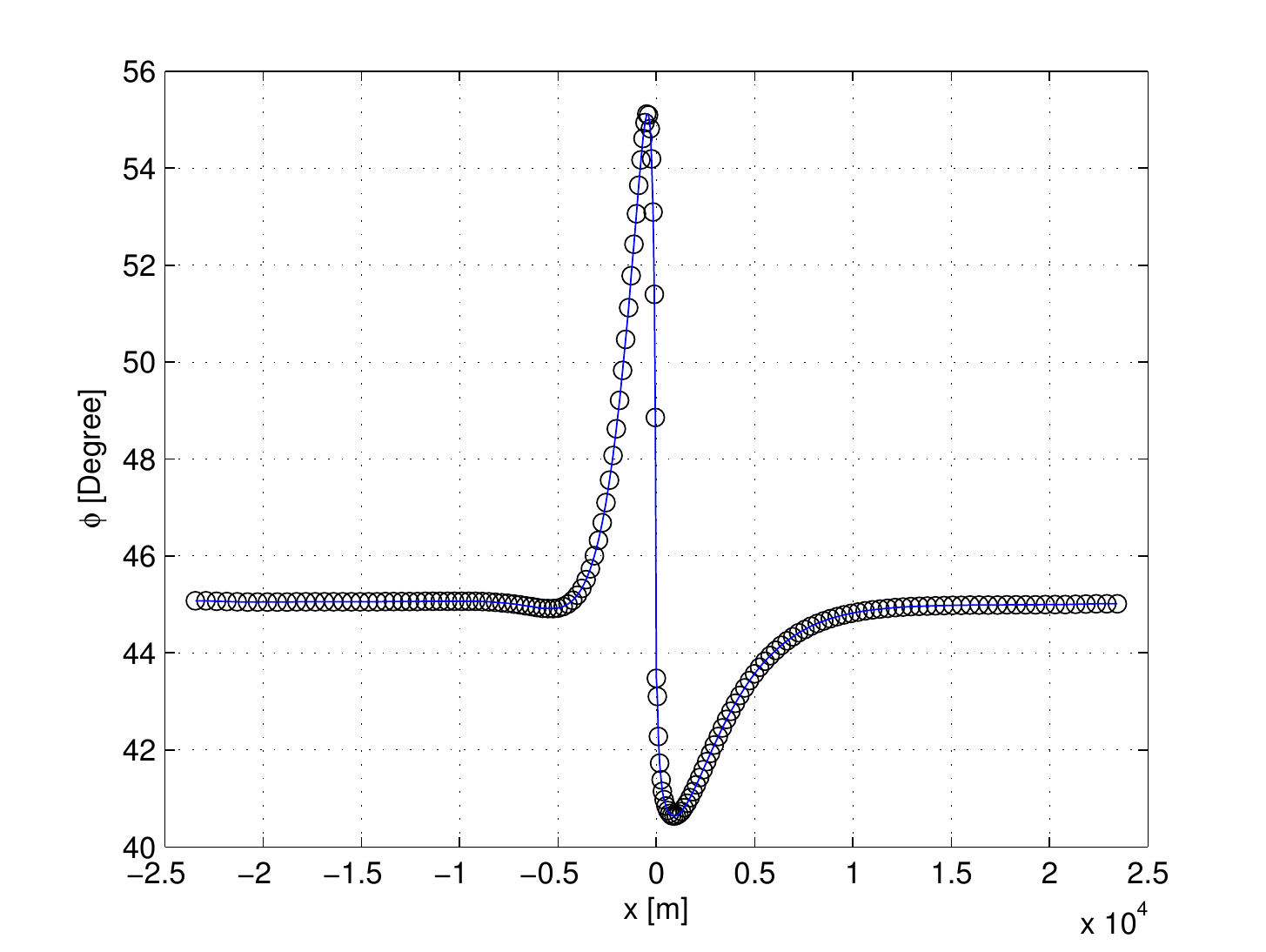}
    \end{subfigure}
    \caption{Phases for the TE-mode (left) and the TM-mode (right) for the quarter-space model using $f=1\,\textup{Hz}$.}
    \label{fig:QuarterSpacePhases}
\end{figure}

The apparent resistivities and phases for the quarter-space model are shown in Fig.~\ref{fig:QuarterSpaceResistivities} and~\ref{fig:QuarterSpacePhases}, respectively. They align closely with the results presented in~\cite{fisc92a,fisc93a}.

\subsection{Rectangular block in half-space solution}

\begin{figure}[!ht]
  \centering
  \includegraphics[width=0.5\textwidth]{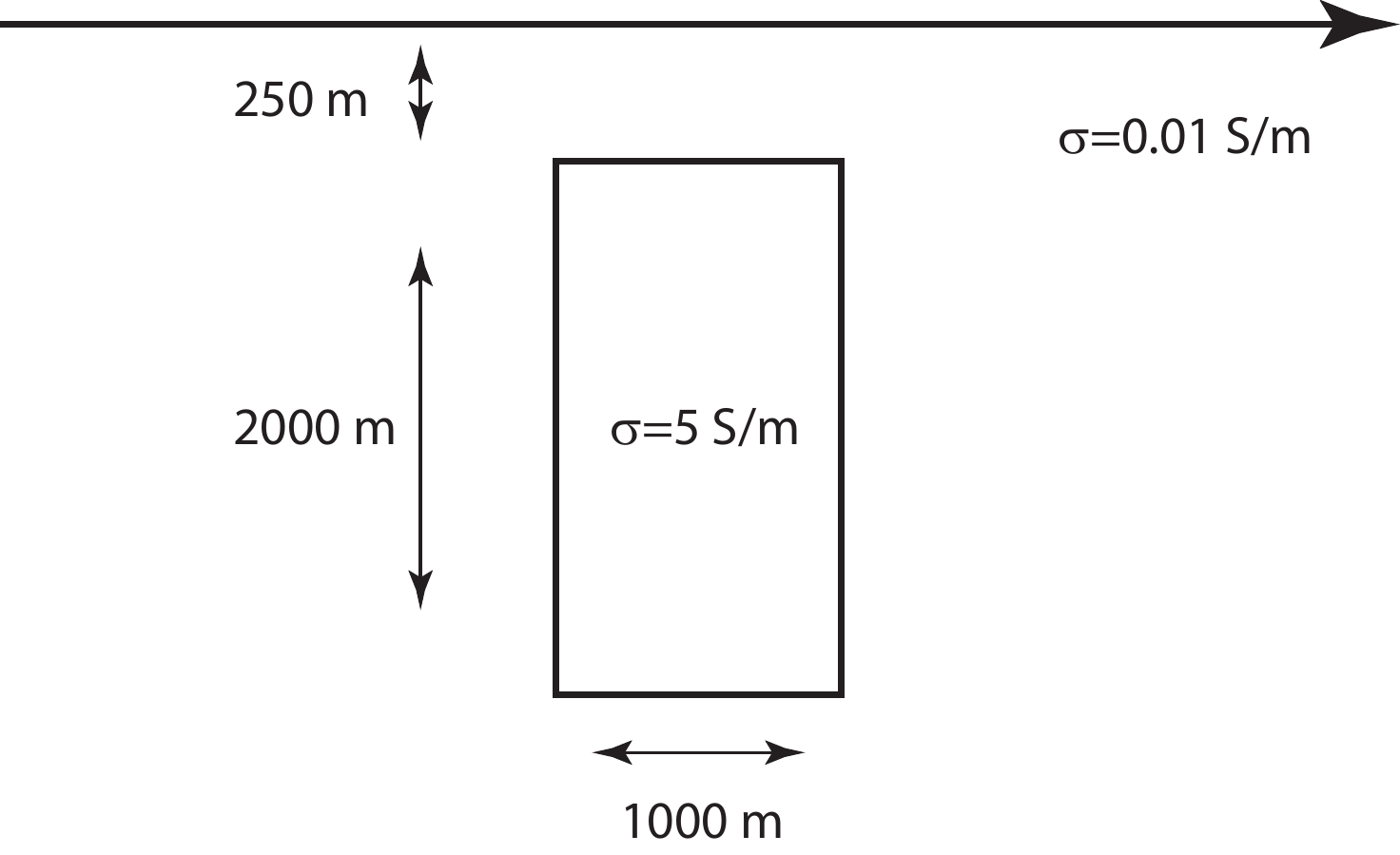}\\
  \caption{Conductivity model for the COMMEMI 2D-1 experiment.}
  \label{fig:BlockInHalfSpaceConductivityModel}
\end{figure}

This experiment coincides with the COMMEMI 2D-1 example~\cite{zhda97a}. It is given by a symmetrical, rectangular, highly conducting block embedded in an otherwise uniform conducting half-space. More precisely, the rectangular block measures $1000\, \textup{m}$  in $x$-direction, $2000\, \textup{m}$ in $z$-direction, with its top edge lying at $z=250\, \textup{m}$. The conductivity of the block is $\sigma=2\, S/\textup{m}$, and the conductivity of the half-space is $\sigma=0.01\, S/\textup{m}$. The conductivity model of this test problem is depicted in Fig.~\ref{fig:BlockInHalfSpaceConductivityModel}. The frequency used in the experiments was $f=10\,\textup{Hz}$. We carry out two experiments for the COMMEMI 2D-1 model.

In the first experiment we obtain a solution to the two-dimensional Maxwell's equations using the stochastic domain decomposition algorithm. For this experiment, the grid cells of the model were of size $\Delta x\times\Delta z=100\,\textup{m}\times 125\,\textup{m}$ throughout the entire domain. The number of Monte-Carlo simulations used in the stochastic solver was $N=5000$.

As was outlined in Section~\ref{sec:SDDForMaxwellsEquations}, the stochastic solution to Maxwell's equations allows one to compute the solution at single points only. For the sake of demonstration, in the second experiment we compute the solution stochastically only in the COMMEMI locations, $x\in\{0,500,1000,2000,4000\}$. More specifically, we compute the solution in three points near the surface at the aforementioned $x$-locations to be able to compute the required secondary fields by evaluating Eqs.~\eqref{eq:MaxwellsEquationSecondary} using regular centered differences. A total of $N=400 000$ Monte-Carlo simulations was used in this experiment. This high number of Monte-Carlo simulations ensures that the primary fields~$E^y$ and~$H^y$ are computed with high accuracy to then allow generating sufficiently accurate approximations for the secondary fields~$E^x$ and~$H^x$, yielding accurate values for the apparent resistivities~$\rho_a^{\rm TE}$ and~$\rho_a^{\rm TM}$.

\begin{figure}[!ht]
    \centering
    \begin{subfigure}[b]{0.5\textwidth}
        \centering
        \includegraphics[width=\textwidth]{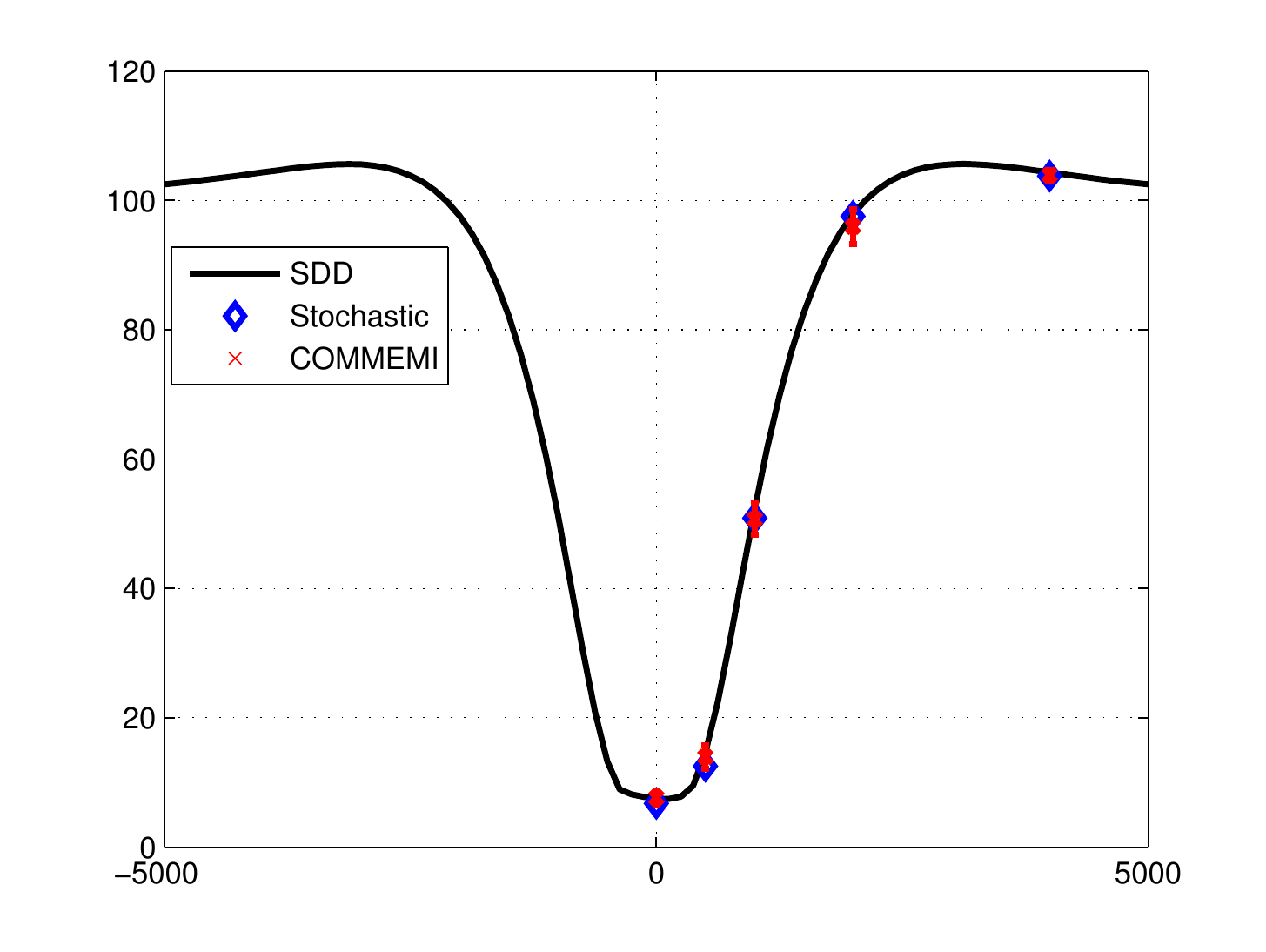}
    \end{subfigure}~
    \begin{subfigure}[b]{0.5\textwidth}
        \centering
        \includegraphics[width=\textwidth]{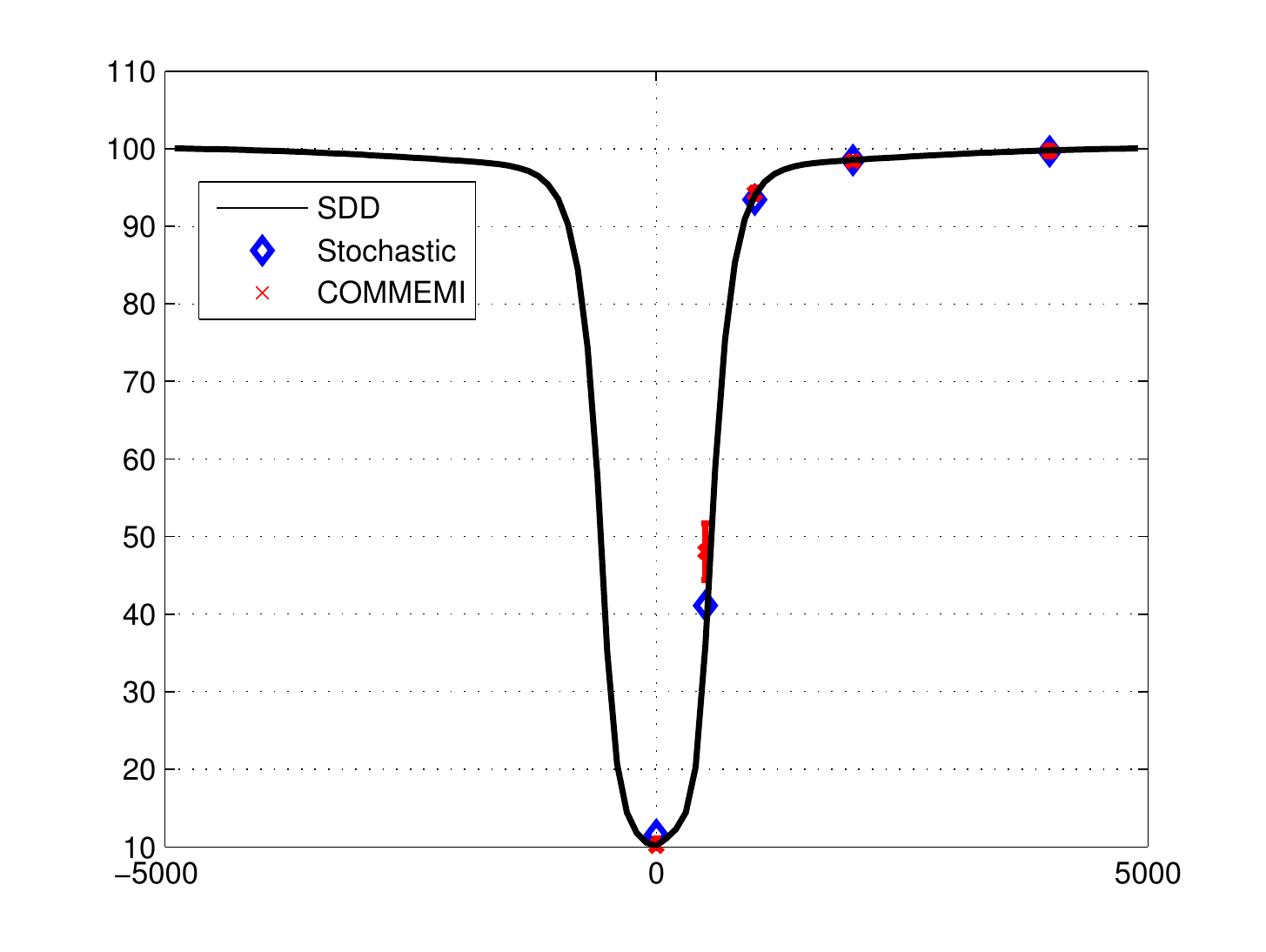}
    \end{subfigure}
    \caption{Apparent resistivities for the TE-mode (left) and the TM-mode (right) for the COMMEMI 2D-1 experiment using $f=10\,\textup{Hz}$.}
    \label{fig:Commemi2DResistivities10Hz}
\end{figure}

The apparent resistivities for the TE-mode and TM-mode are shown in Fig.~\ref{fig:Commemi2DResistivities10Hz}.

To give a better comparison with the values reported in the COMMEMI experiments, in Table~\ref{tab:COMMEMIComparison} we list the mean values (and standard deviation) taken from Table~B.8 in~\cite{zhda97a} along with the numerical values obtained with our two approaches.

\begin{table}[!ht]
\caption{Apparent resistivities computed using the SDD method and the purely stochastic algorithm compared to the original COMMEMI results.}
\begin{tabular}{cccccc}
\hline
\hline
$\rho_a\textup{(TM)}$ & 0 m & 500 m & 1000 m & 2000 m & 4000 m\\
\hline
SDD & 10.15 & 36.01 & 93.98 & 98.55 & 99.78 \\
Stochastic & 11.58 & 41.10 & 93.43 & 98.52 & 99.65 \\
COMMEMI & 10.13 $\pm$ 0.96 & 48.07 $\pm$ 3.65 & 94.27 $\pm$ 0.79 & 98.40 $\pm$ 0.40 & 99.71 $\pm$ 0.64\\
\hline
\hline
$\rho_a\textup{(TE)}$ & 0 m & 500 m & 1000 m & 2000 m & 4000 m\\
\hline
SDD & 7.44 & 13.27 & 51.20 & 97.58 & 104.36 \\
Stochastic & 6.70 & 12.50 & 50.84 & 97.54 & 103.77 \\
COMMEMI & 7.60 $\pm$ 1.04 & 13.92 $\pm$ 1.82 & 50.70 $\pm$ 2.48 & 95.94 $\pm$ 2.75 & 103.92 $\pm$ 0.80\\
\hline
\hline
\end{tabular}
\label{tab:COMMEMIComparison}
\end{table}

It can be seen from Table~\ref{tab:COMMEMIComparison} that the stochastic domain decomposition method produces values that are well within the range of results reported in the COMMEMI experiments. The only significant deviation is the value for the TM-mode resistivity at $x=500\,\textup{m}$. As can be seen from the right plot in Fig.~\ref{fig:Commemi2DResistivities10Hz}, this is the region of highest variability in the resistivity and the COMMEMI mean is obtained between $x=500\,\textup{m}$ and the neighboring grid point. Similarly, the point-wise solution obtained using the purely stochastic algorithm also gives results that are well within the range of the COMMEMI results, demonstrating that if solutions are sought in single points only, the stochastic algorithm may be a viable alternative compared to standard deterministic methods that requires the computation of the numerical solution over the entire domain even if the solution is required at several points only.

\subsection{Triangular block in half-space solution}

\begin{figure}[!ht]
  \centering
  \includegraphics[width=0.5\textwidth]{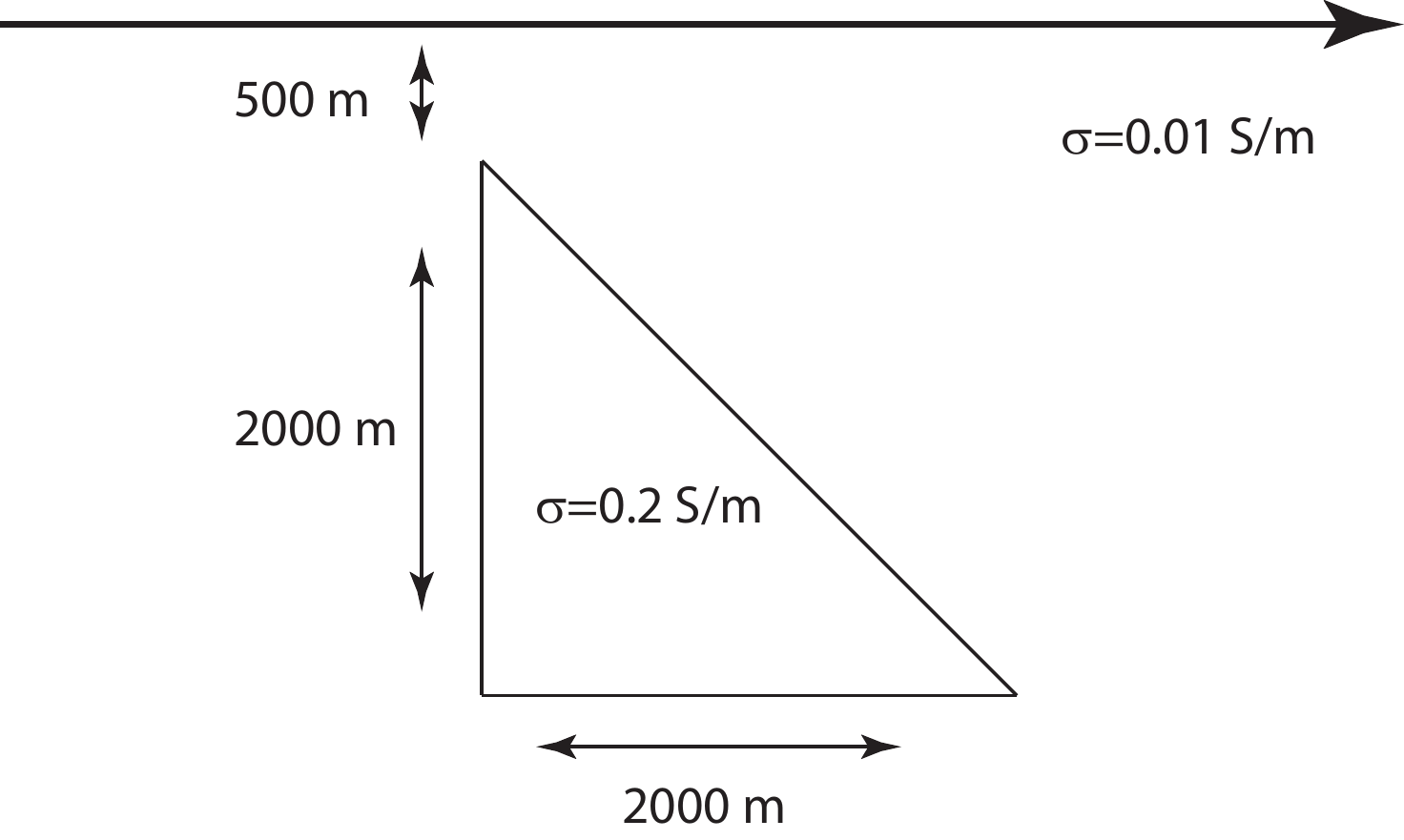}\\
  \caption{Conductivity model for the triangle in a half-space example.}
  \label{fig:TriangleInHalfSpaceConductivityModel}
\end{figure}

This experiment was previously considered in~\cite{farq07a}. It is a bit more general than the COMMEMI 2D-1 example and, with the sloping interface of the triangular anomaly, begins to illustrate the capability of the combination of the domain-decomposition solver and the meshless sub-domain solver to take into account arbitrary, complex interfaces. The conductivity model for this example is illustrated in Fig.~\ref{fig:TriangleInHalfSpaceConductivityModel}. The triangle has corners at the three points $(-600,400)$, $(-600,2500)$ and $(1500,2500)$ with conductivity $\sigma=0.2\, S/\textup{m}$ in a half-space with conductivity $\sigma=0.01\, S/\textup{m}$.

The grid cells for this model were of size $\Delta x\times\Delta z=100\,\textup{m}\times 50\,\textup{m}$ and $N=5000$ Monte-Carlo simulations were used for the approximation of the expected values. For this experiment, we used the frequencies $f=1\,\textup{Hz}$, $f=3\,\textup{Hz}$ and $f=10\,\textup{Hz}$.

\begin{figure}[!ht]
    \centering
    \begin{subfigure}[b]{0.5\textwidth}
        \centering
        \includegraphics[width=\textwidth]{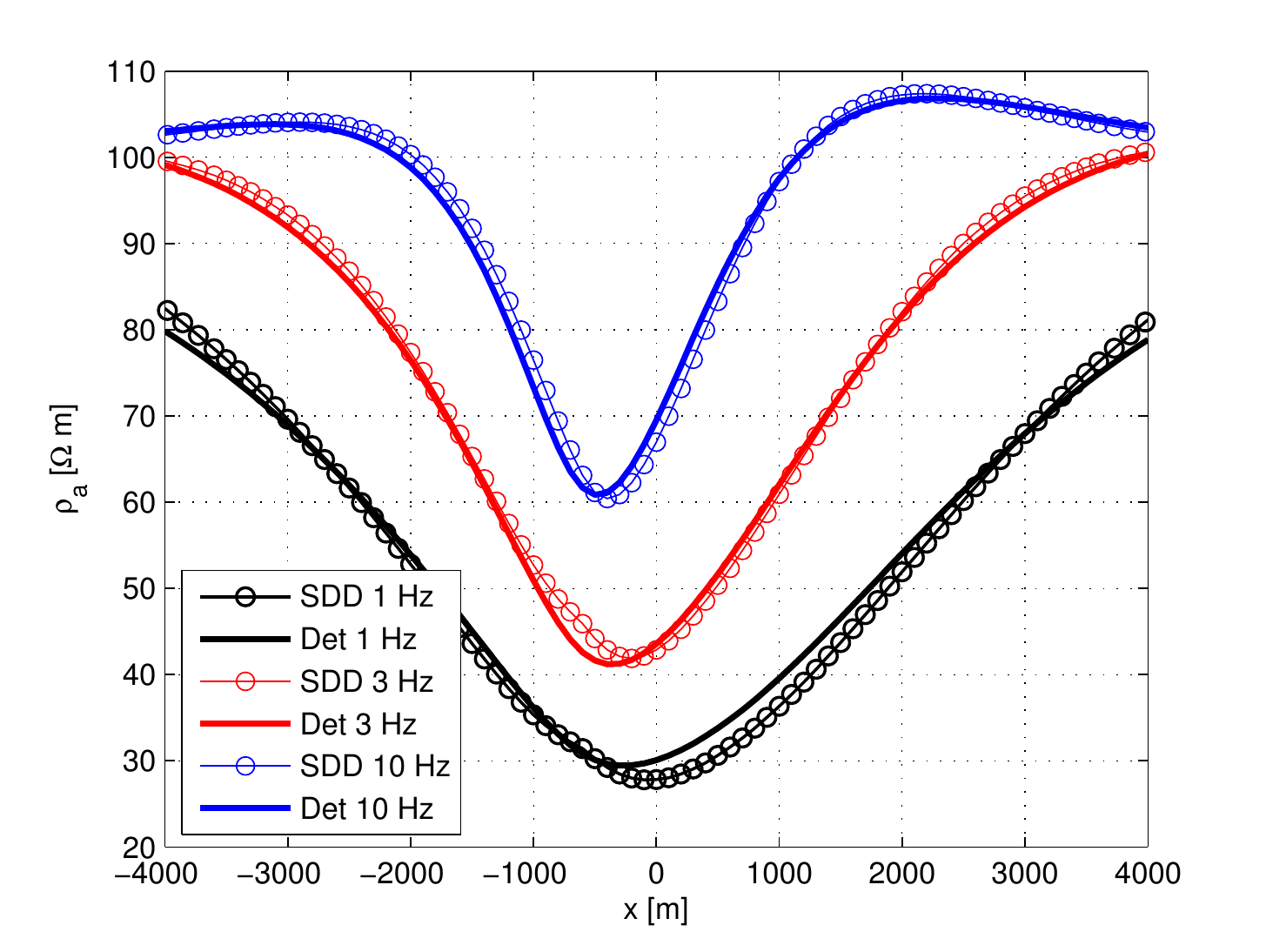}
    \end{subfigure}~
    \begin{subfigure}[b]{0.5\textwidth}
        \centering
        \includegraphics[width=\textwidth]{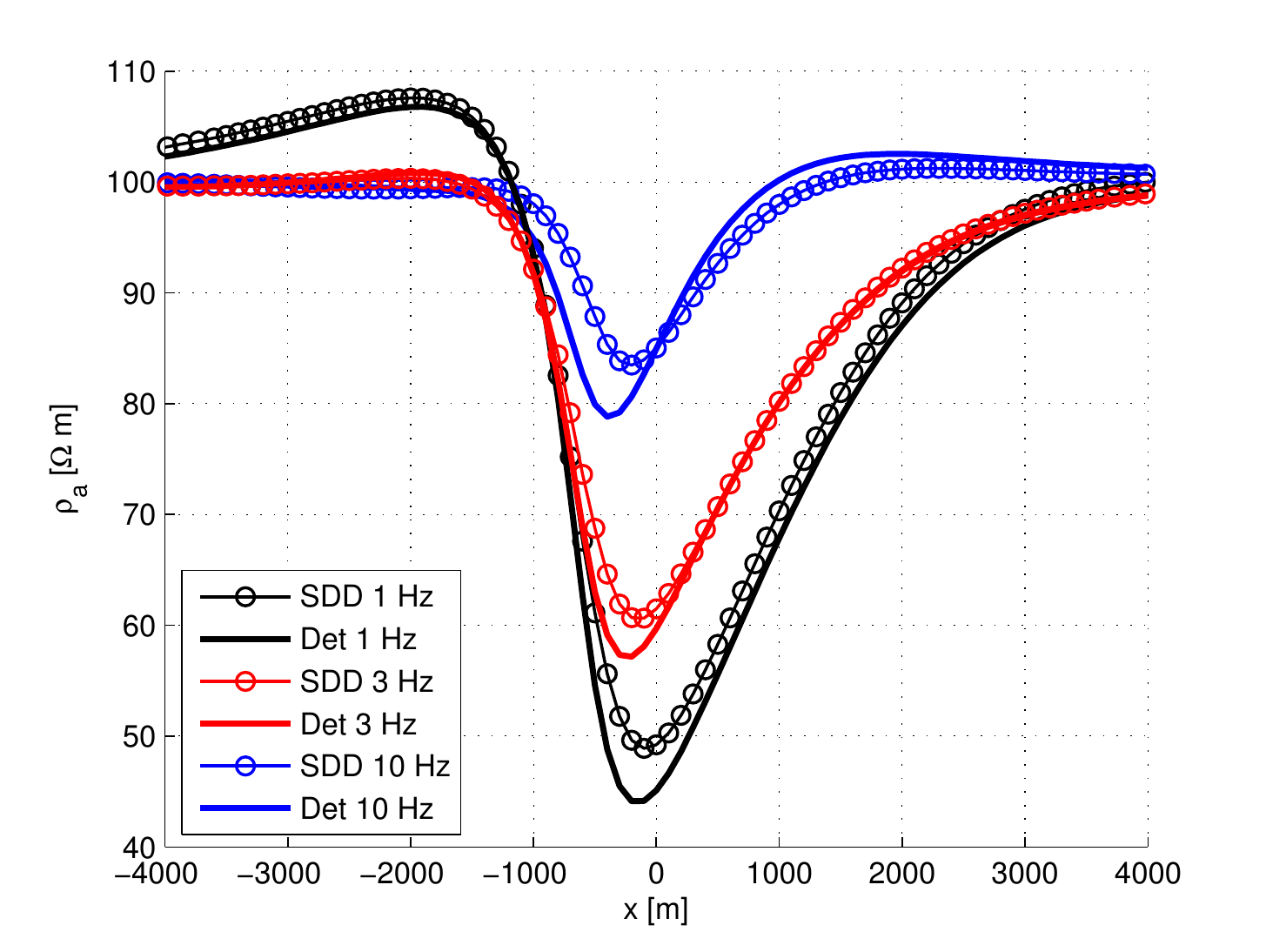}
    \end{subfigure}
    \caption{Apparent resistivities for the TE-mode (left) and the TM-mode (right) for the triangle in a half-space experiment. Results from the SDD model (circles) and the deterministic model presented in~\cite{farq07a} for the frequencies $f=1\, \textup{Hz}$, $f=3\,\textup{Hz}$ and $f=10\,\textup{Hz}$.}
    \label{fig:TriangleResistivitiesComparison}
\end{figure}

\begin{figure}[!ht]
    \centering
    \begin{subfigure}[b]{0.5\textwidth}
        \centering
        \includegraphics[width=\textwidth]{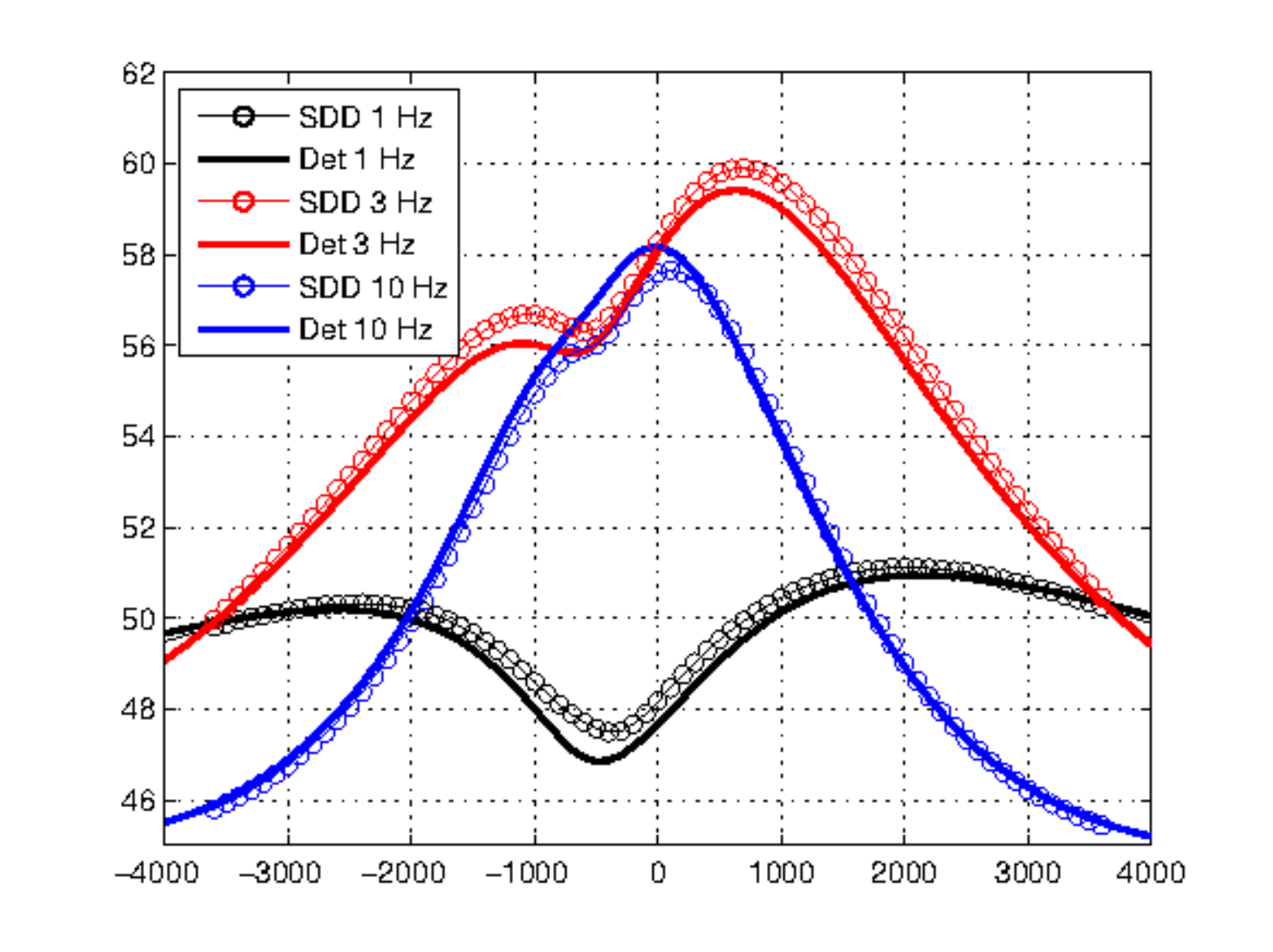}
    \end{subfigure}~
    \begin{subfigure}[b]{0.5\textwidth}
        \centering
        \includegraphics[width=\textwidth]{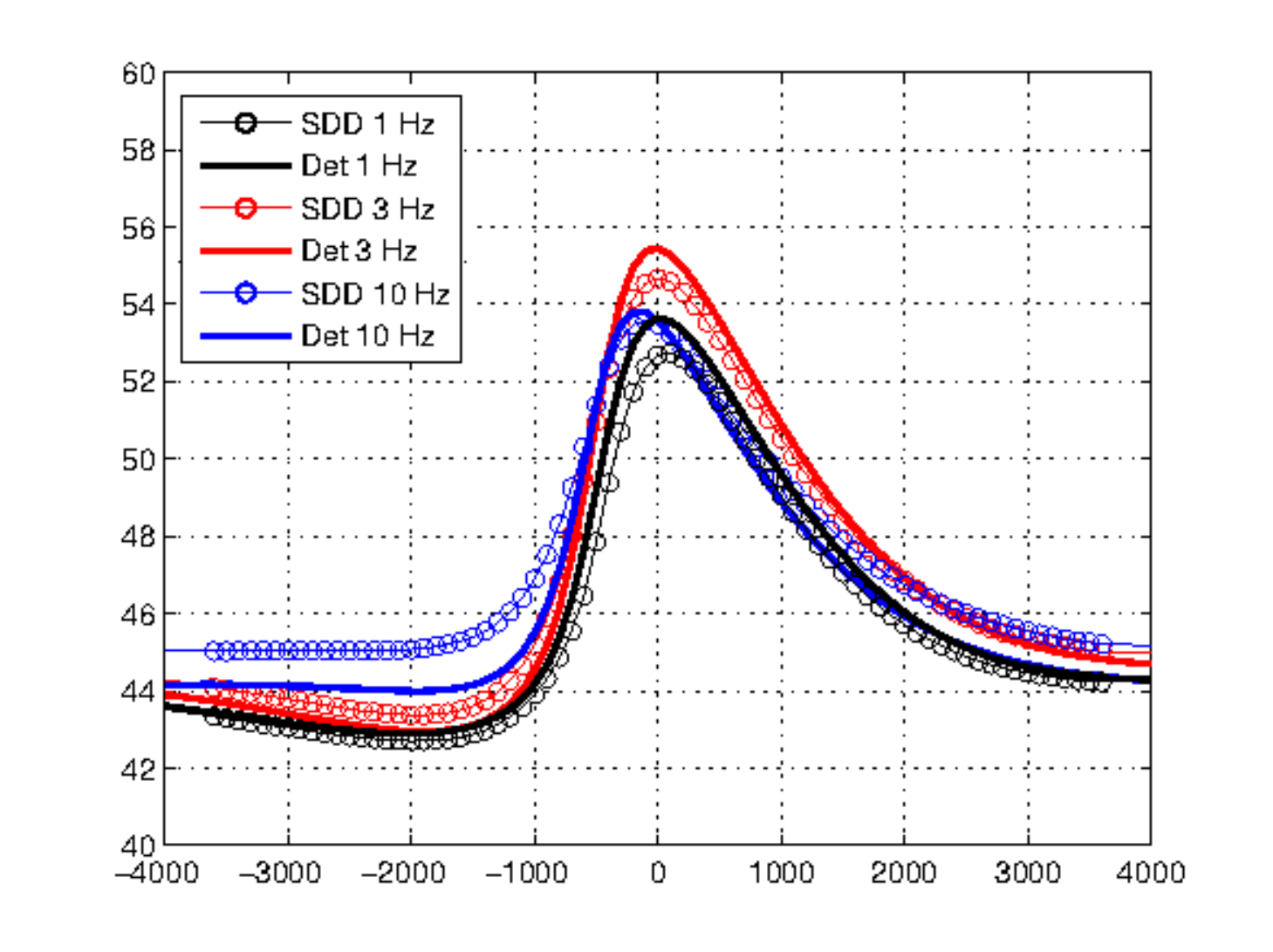}
    \end{subfigure}
    \caption{Phases for the TE-mode (left) and the TM-mode (right) for the triangle in a half-space experiment. Results from the SDD model (circles) and the model presented in~\cite{farq07a} for the frequencies $f=1\, \textup{Hz}$, $f=3\,\textup{Hz}$ and $f=10\,\textup{Hz}$.}
    \label{fig:TrianglePhasesComparison}
\end{figure}

The conductivities and phases for this experiment are displayed in Fig.~\ref{fig:TriangleResistivitiesComparison} and Fig.~\ref{fig:TrianglePhasesComparison}. Here we present the results using the SDD method and the model developed in~\cite{farq07a}.

It can be seen from Fig.~\ref{fig:TriangleResistivitiesComparison} and Fig.~\ref{fig:TrianglePhasesComparison} that for the TE-mode the apparent resistivities and phases for both methods coincide closely. For the TM-mode the results do not coincide as well, with the discrepancy increasing as the frequency increases. The reason for this is that the conductivity model is treated differently by the different methods. For the stochastic domain decomposition approach presented here, a conductivity is associated with each node, with this conductivity being implicitly an average over the neighbourhood of the node. For the FD scheme of~\cite{farq07a}, the conductivity is explicitly considered to be uniform throughout each rectangular cell of the mesh with the approximate values for $E_x$ and $H_x$ solved for at cell centers and cell vertices (for the TE- and TM-modes respectively).

\section{Conclusion}\label{sec:ConlusionSDDMaxwellsEquations}

The present paper introduced the stochastic domain decomposition method for solving the two-dimensional Maxwell's equations as required in the magnetotelluric method. The method is new in that it allows splitting of the sub-surface into regions of constant or continuous conductivity, over which Maxwell's equations can be solved independently. This splitting also allows one to use the strong form of Maxwell's equations and thus the potential costly numerical integrations required in solvers using the weak form can be avoided. The interface solutions for these sub-domains are naturally found by evaluating the stochastic form of Maxwell's equations numerically using Monte-Carlo techniques. Once these interface values have been computed, any sub-domain solver can be used to obtain the solution over the entire physical domain. Here we have used a deterministic sub-domain solver based on radial basis function based finite differences. We argue that such a solver is suitable for magnetotelluric modeling as it allows one to work with irregularly shaped sub-domains, which arise naturally in realistic sub-surface models. 

While Monte-Carlo methods are notoriously costly, invoking them only within the framework of stochastic domain decomposition makes for an efficient way of solving partial differential equations, particularly if massively parallel computing architectures are available. Since these architectures are getting more and more popular, stochastic domain decomposition becomes an attractive alternative to conventional parallelization methods. We also note here that the single interface values can be computed independently of each other which is essential for the parallelization of the algorithm. The computational benefits of stochastic domain decomposition where already established in several scaling studies, see e.g.~\cite{aceb05a,aceb10a,bihl14a}.

Moreover, there are several possibilities for accelerating the computation of the stochastic part of the problem, such as computing the stochastic solution only in certain points along the interface and using interpolation to obtain the remaining interface values. This procedure has proved successful in the application of the stochastic domain decomposition method to both solving physical PDEs~\cite{aceb05a} and generating adaptive moving meshes~\cite{bihl14a}. Further speed-up can be obtained by using GPU computing for the solution of the stochastic differential equations, see e.g.~\cite{prei09a,vanM08a} for some examples. These avenues will be explored in a forthcoming work.

We should again like to stress that while the bulk of this paper was devoted to the idea of evaluating the stochastic form of the exact solution of Maxwell's equations to obtain interface values separating regions of constant conductivity, the point-wise nature of this solution also allows one to compute the solution at specific points only. This can be of interest if the solution to the magnetotelluric problem is only required near measurement sites. As a demonstration of this property, we computed the solution for the block-in-half-space example (the COMMEMI 2D-1 example) only at regional key points. This property can be attractive if a solution is sought in distinct points over a large domain, since it bypasses the need to obtain the solution over the entire domain as required in traditional deterministic methods.

The examples studied in the present paper are quite simple. They should be regarded as a proof of the concept and to demonstrate that stochastic domain decomposition is a viable alternative to more traditional ways of discretizing Maxwell's equations. More realistic sub-surface models are under investigation and will be the subject of a future paper.

\section*{Acknowledgements}

This research was undertaken, in part, thanks to funding from the Canada Research Chairs program (AB) and the NSERC Discovery Grant Program (CGF,RDH,JCLO). The authors thank Antoine Lejay (INRIA), Scott MacLachlan (MUN) and Paul Tupper (SFU) for helpful discussions.

{\footnotesize\setlength{\itemsep}{0ex}
 \bibliography{bihlo}
}

\end{document}